\documentclass[a4paper]{article}
\usepackage[latin1]{inputenc} %
\usepackage[T1]{fontenc} %
\usepackage{RRA4}
\usepackage{amsmath,amssymb}
\usepackage{subfigure}
\usepackage{hyperref}
\RRdate{September 2009}

\RRauthor{
Fabien Mathieu\thanks[orange]{Orange Labs}%
  \and
Diego Perino\thanks{INRIA} \thanksref{orange}%
}
\authorhead{Mathieu \& Perino}
\RRtitle{Algorithmes de diffusion des ressources avec gestion des capacités}
\RRetitle{On Resource Aware Algorithms\\ in Epidemic Live Streaming}
\titlehead{On Resource Aware Algorithms in Epidemic Live Streaming}
\RRresume{Les algorithmes de diffusion épidémique sont une des solutions possibles pour faire de la diffusion multimedia en quasi-direct sur des réseaux pair-à-pair. Leur efficacité a été largement étudiée dans les réseaux homogènes, où tous les pairs ont la même bande passante disponible. En revanche, la diffusion épidémique dans un système hétérogène est encore peu comprise.

En se focalisant sur le mécanisme de sélection des destinataires, nous proposons un modèle générique qui englobe une large classe d'algorithme pour réseaux hétérogènes. La sélection est décomposée en deux fonctions, dont l'une tient compte du réseau sous-jacent tandis que l'autre est agnostique.

Au travers de simulations, nous analysons le compromis entre adaptation au réseau et agnostisme dans des réseaux hétérogènes. Nous mettons en évidence l'importance de la bonne diffusion des toutes premières copies d'un chunk donné, ainsi que la possibilité de régler l'équité du système entre pairs riches et moins riches.}
\RRabstract{Epidemic-style diffusion schemes have been previously proposed for
achieving peer-to-peer live streaming.
Their performance trade-offs have been deeply analyzed for homogeneous
systems, where all peers have the same upload capacity. However,
epidemic schemes designed for heterogeneous systems have not been
completely understood yet.

In this paper we focus on the peer selection process and propose a
generic model that encompasses a large class of algorithms.
The process is modeled as a combination of two functions, an aware one and an agnostic one.

By means of simulations, we analyze the awareness-agnostism trade-offs
on the peer selection process and the impact of the source
distribution policy in non-homogeneous networks.
We highlight  that the early diffusion of a given chunk is crucial for
its overall diffusion performance, and a fairness trade-off arises between the performance of heterogeneous peers, as a function of the level of awareness.}
\RRmotcle{Pair-à-pair, diffusion épidémique, hétérogénéité, équité}
\RRkeyword{P2P, Epidemic Live Streaming, Heterogeneous bandwidth, Fairness}
\RRprojet{Gang}  
\RRdomaine{1} 
\RRtheme{COM -- Systèmes communicants}
\RRNo{7031}
\URRocq 

\newcommand{\T}{{\cal T}}

\begin{document}
\makeRR   

\tableofcontents
\newpage

\section{Introduction}
Live streaming over the Internet has become increasingly popular in the last few years. To support large audiences that grow over time, the peer-to-peer approach has been proposed by several commercial systems that are now widely used like PPLive~\cite{pplive:http}, SopCast~\cite{sopcast:http}, TVants~\cite{tvants:http} and UUSee~\cite{uusee:http}. These systems rely on unstructured, chunk-based diffusion algorithms: the stream is divided in a series of pieces (chunks), that are injected in the system by the source and exchanged among peers in order to retrieve the complete sequence and play out the stream.

The theoretical performance trade-offs of such chunk-based systems have been deeply analyzed for homogeneous scenarios, where all peers have the same upload capacity. However, most peer-to-peer systems are heterogeneous by nature, and the impact of that heterogeneity has not been completely understood yet.

This paper aims at clarifying the handling of heterogeneity for epidemic-style diffusion algorithms, where the chunk exchanges are mainly decided at senders' side (push approach). We propose to give a
generic model that encompasses a large class of algorithms, and to discuss some results and experiments based on that model.

\subsection{Related Work}

Chunk dissemination algorithms are hard to analyze because of the strong interaction imposed by the chunk exchanges. The exchange algorithms run locally at every node, and can be described by chunk/peer selection policies. Although the local policies can be very simple, the whole network often behaves as a complex system, making the study of its performance complicated.
However, analytical results  have been derived for homogeneous systems where peers all have the same upload capacity. Schemes achieving optimal diffusion rate are analyzed in~\cite{zhang.xiong.ea:optimizing,zhang.zhang.ea:understanding,massoulie.twigg.ea:randomized}. A scheme that  achieves optimal diffusion delay is proposed in~\cite{sanghavi.hajek.ea:gossiping}, while algorithms providing optimal diffusion rate within an optimal delay are studied in~\cite{sigepidemic,trentoepidemics}. Performance trade-offs of epidemic-style algorithms are deeply analyzed for homogeneous systems in~\cite{sigepidemic,gyorgy:delay}.

In heterogeneous systems, where peers have different upload capacities, dissemination algorithms should take into account the capacities of the nodes somehow, in order to improve the performance, but a certain  level of altruism is required for the functioning of the system. In other words, a kind of equilibrium should be found that ensures a good utilization of the powerful nodes, while guarantying that weaker nodes are not excluded from the diffusion process. Live streaming diffusion schemes that aim at finding such an equilibrium have been proposed and analyzed by means of simulations~\cite{silva.leonardi.ea:bandwidth-aware,ross:p2ptv} or experimental evaluations~\cite{picconi.massoulie:is,pulseawareness}.

Analytical studies of resource aware unstructured algorithms for P2P systems have mainly been performed for file-sharing~\cite{qiu.d.srikant.r:modeling,gai.a.mathieu.f.ea:stratification}, or for generic applications by means of a game theory approach~\cite{chiranjeeb:game,ma:incentive,chao.lui:mathematical}. 
As concern live streaming, Chu et al.~\cite{chu.zhang:considering} propose a framework to evaluate the achievable download performance of receivers as a function of the altruism from the bandwidth budget perspective. They highlight that altruism has a strong impact on the performance bounds of receivers and that even a small degree of altruism brings significant benefit. 
In~\cite{lin:game} a game-theoretic framework is proposed to model and evaluate incentive-based strategies to stimulate user cooperation.

\subsection{Contribution}

In Section~\ref{sec:epidemics_aware_model}, we propose a model for unstructured P2P live streaming diffusion schemes that takes explicitly the awareness-agnostism trade-off into account. This model is highly versatile, so it can represent several existing resource-aware peer selection policies, as well as new ones. Then in Section~\ref{sec:hete_diff} we propose recursive formulas for the diffusion function of a generic \emph{resource aware peer/latest blind chunk} selection scheme. Lastly, by means of simulations, we deeply analyze in Section~\ref{sec:simus} the awareness-agnostic trade-off and the critical role the source policy plays in the system performance.

\section{Model and schemes}
\label{sec:epidemics_aware_model}

We consider a P2P system of $n$ peers receiving a live stream from a single source $S$. We suppose that peers have a partial knowledge of the overall system that is represented by an Erd{\"o}s-Renyi $\mathcal{G}(n+1,p_e)$ graph (the source has a partial knowledge of the system like any other peer). We denote the set of neighbors of peer $l$ as $N(l)$ and we suppose a peer can only send chunks to one of its neighbors. 

We suppose that every peer $l$ has a limited upload capacity $u(l)$ and that there is no constraint on the quantity of data that each peer can receive per time unit. For simplicity, we assume that the bandwidth distribution is discrete, with $U$ possible distinct values, and we partition the peers in $U$ classes $C_1$, \ldots $C_U$ according to their upload capacity. We denote as $\alpha_i$ the percentage of peers belonging to class $C_i$. The source has a limited upload capacity as well, denoted as $u_S$.

We suppose that the stream has a constant rate $SR$. The source splits it in a sequence of chunks of size $c$, so that a new chunk is created every $T_{SR}=\frac{c}{SR}$ time units. These chunks are injected into the system according to the source diffusion policy and upload constraints. The peers in turn exchange these chunks among them according to their own diffusion policy, which may differ from the one of the source. For every peer $l$, let $B(l)$ be the collection of chunks that peer $l$ has received. 


A convenient way to represent a diffusion policy is to decompose it in a peer selection process and a chunk selection process, which can be performed in the \emph{peer-then-chunk} or in the \emph{chunk-then-peer} order.

In this paper, we limit ourselves to diffusion schemes where the peer is selected first, although the model presented could be extended to the \emph{chunk-then-peer} case. We argue that if the chunk is selected first, the peer selection is restricted to the peers missing the given chunk, so that resource awareness is potentially less effective. Moreover, peer-first schemes have been shown more adapted to a practical implementation because they potentially generate low overhead and provide near-optimal rate/delay performance, while chunk-first schemes tend to generate a lot of signaling messages~\cite{sigepidemic}.

Regarding the selection processes themselves, we focus here on the peer selection process, while for the chunk selection we just consider two simple policies called \emph{latest blind (LB)} and \emph{latest useful (LU)}, which have been shown efficient in homogeneous environments~\cite{sigepidemic}.
If a peer runs a \emph{latest blind} chunk policy, it sends to the selected peer the more recent chunk generated by the source it owns. This minimizes the need for communication between peers, but increases the chances of wasting bandwidth by sending a chunk already received by the destination. On the other hand, with the \emph{latest useful} chunk policy, a peer sends to the receiver peer the more recent chunk it owns that the receiver peer has not downloaded yet, if any. This requires at least one message exchange between the two peers. In both cases (blind or useful), the sending time of peer $l$ of class $i$ is defined by $T_i=\frac{c}{u_i}$ if the selected chunk is indeed useful for the destination peer. If not, the destination peer can send back a notification so that the sender can select another peer.

The reason why we only consider these two simple chunk policies is that we believe that chunk selection is less crucial than peer selection for heterogeneous peers. Of course, this is true only if chunks are all equal in size and if they all have the same importance: if some chunks have higher priority or are bigger than others, for example because they have been coded with layered techniques, the chunk selection policy play an important role~\cite{ross:p2ptv}. However the study of chunk-differentiated scenarios is beyond the scope of this paper, so we focus on the impact of the peer selection process.

\subsection{Peer Selection Process}

We now propose a general model that allows to represent various non-uniform peer selection schemes.
The non-uniform selection is represented by weight functions $\{H_l\}$. A peer $l$ associates to every neighbor $v \in N(l)$ a weight $H_l(v)$. Typical weight functions will be expressed later for some schemes. $H_l(v)$ can be time-dependent, however the time variable is implicit in order not to clutter notation. 

Whenever a given peer $l$ can upload a chunk, we assume it can use one of the two following peer selection policies:
\begin{itemize}
	\item \textbf{Aware}  peer $l$ selects one of its neighbors $v \in N(l)$ proportionally to its weight $H_l(v)$.
	\item \textbf{Agnostic} peer $l$ selects one of its neighbors $v \in N(l)$ uniformly at random.
\end{itemize}

The choice between the two policies is performed at random every time a chunk is sent by a peer, the aware policy been selected with a probability $W$, called the \emph{awareness probability} ($0\leq W\leq 1$).
$W$ expresses how much a peer takes resources into account when performing the selection so that it represents the level of awareness of the diffusion scheme.

The $H_l$ function and the $W$ variable completely define the peer selection scheme: when a peer $l$ can upload a chunk, the probability $\beta(l,v)$ that it selects one of its neighbors $v$ is therefore given by

\begin{equation}
\beta(l,v)=\underbrace{\frac{H_l(v)}{\sum_{k \in N(l)}H_l(k)}W}_{\text{Aware}}+ \underbrace{\frac{1-W}{N(l)}}_{\text{Agnostic}}
\label{eq:generic_aware_agnostic}
\end{equation}







In the following we express $H$ and/or $W$ for some peer selection schemes. Remember that we consider diffusion schemes where the peer is selected first. This means that, unless otherwise specified, a sender peer has no prior knowledge about the buffer state of its neighbors, so it is not guaranteed that it will have useful chunks for the peer it will select.

\paragraph{Random peer selection (RP)}
The random peer selection is the limit case where peers are completely unaware of their neighbors' characteristics. We then have $W=0$, and there is not need to define a weight function. This results in 
\[\beta(l,v)=\frac{1}{N(l)}\text{.}\]

\paragraph{Bandwidth-aware peer selection (BA)}
This is the simplest scheme taking into account the upload capacities of the nodes. A peer $l$ selects one of its neighbors $v \in N(l)$ proportionally to its upload capacity, so we have $H_l(v) = u(v)$. Note that in the homogeneous upload capacity case, the selection is indeed equivalent to the uniformly random selection.

The bandwidth-aware scheme has been introduced by da Silva \emph{et al.} in~\cite{silva.leonardi.ea:bandwidth-aware}. However there are two main differences between our model and the framework they propose: in \cite{silva.leonardi.ea:bandwidth-aware},
\begin{itemize}
	\item the chunk is selected first, and the bandwidth-aware selection is performed among the neighbors that need the selected chunk from the sender;
	\item the selection scheme is fully-aware (corresponding to $W=1$ in our model), while we propose to discuss later the influence of the awareness probability $W$.
\end{itemize}

Although this paper focuses on a edge-constraint scenario, the upload estimation may differ in practice depending on the measurement points. Our model could be easily generalized by setting $H_l(v)=u_l(v)$, where $u_l(v)$ is the available bandwidth capacity from $v$ to $l$.

 

\paragraph{Tit-for-Tat peer selection (TFT)}

Tit-for-tat mechanisms have been introduced in P2P by the BitTorrent protocol~\cite{cohen.b:incentives}, and have been widely studied for file sharing systems. Such incentive mechanisms can be very effective in live streaming applications~\cite{pulseawareness}.

In the original BitTorrent protocol, a subset of potential receivers is periodically selected~\cite{cohen.b:incentives}. Following the authors in~\cite{ross:p2ptv}, we propose a simpler protocol where a receiver peer is selected every time a chunk is sent. 
We propose to drive the peer selection by using as weight function $H_l(v)$ an historic variable that is computed every \emph{epoch $T_e$};  this historic value indicates the amount of data peer $l$ downloaded from peer $v$ during the last epoch. In this way, a peer $v$ is selected by a peer $l$ proportionally to the amount of data it provided to $l$ during last epoch.


\paragraph{Data-driven peer selection}
The model we introduced so far is not only able to describe the behavior of resource-aware algorithms, but also to represent diffusion schemes that take into account the collection of chunks $B$ when performing peer selection.

The \emph{most deprived selection} presented for instance in~\cite{sigepidemic}, as well as the \emph{proportional deprived selection} proposed by  Chatzidrossos et al.~\cite{gyorgy:delay}, can be represented by our model.

The former selects the destination peer uniformly at random among those 
neighbors $v$ of $l$ for which  $|B(l) \setminus B(v)|$ is maximum. The weight function can be expressed as:

\begin{equation}
H_l(v)= \left\{
\begin{array}{l}
1\text{ if } |B(l) \setminus B(v)|=\max_{v \in N(l)} |B(l) \setminus B(v)|\text{,}\\
0\text{ otherwise.}
\end{array}
\right.
\label{eq:weight_md}
\end{equation}

%


The latter selects a destination peer $v$ proportionally to the number of useful chunks the sender peer $l$ has for it. The weight function can be expressed as $H_l(v) = |B(l) \setminus B(v)|$.

In the following we are not going to analyze these data-driven peer selection schemes because we focus on resource-aware policies. However, the recursive formulas derived in Section~\ref{sec:epidemics_aware_formulas} are also valid for data-driven peer selection policies.

\subsection{Performance evaluation}

Following \cite{sigepidemic}, we focus on the achieved rate and delay to assess the performance of a given diffusion scheme. In details, we call \emph{rate} the asymptotic probability that a peer (random or belonging to a specific class) receives a given chunk. On the contrary, the \emph{chunk miss ratio} is the asymptotic probability to miss a chunk (or equivalently the difference between the stream rate $SR$ and the actual goodput). 
Note that links are supposed to be lossless, so a peer misses a given chunk only if none of its neighbors has scheduled that chunk for it.
The \emph{average diffusion delay} is defined as the time needed for a chunk to reach a peer on average. For practical reasons, we assume a fixed diffusion deadline: chunk transmissions that occur too long after the chunk's creation are not taken into account; the deadline is by construction an upper bound for the transmission delay. 

For a fully random scheme, one should expect the performance to be roughly the same for all peers, as there is no reason for one peer to be advantaged compared to another. This is not the case for schemes with $W>0$, so we may have to use a per class performance evaluation.

\subsection{Implementation issues}
\label{sec:epidemics_aware_implementation}

The simplicity  and strength of the bandwidth-aware selection comes from the fact that it directly uses the amount of bandwidth provided by a node as weight function. The upload capacity can be measured  by means of bandwidth estimation tools, or can be provided by an external oracle/tracker. However, both approaches highlight several practical drawbacks.

In the case of measurements made by the peers themselves, known bandwidth estimation tools may be inaccurate, particularly when used in large-scale distributed systems~\cite{croce:quest}. Moreover, the measured value may vary over time according to network condition, so that the measurement should be frequently repeated generating high overhead and interference.

If some tracker or oracle is used, the upload capacity monitored by the central authority can be a nominal one, provided by the peers, or can be inferred from measurements made from different points. Apart from accuracy issues, the authority providing the information, as well as the measurement points, should be trusted and should not cheat on the values they provide.

In our model we do not take all these issues into account, but we argue that this scheme is currently hard to implement in real systems. However, some projects, like Napa-Wine~\cite{napa:http}, or standardization efforts, like ALTO~\cite{alto:http}, are working in order to provide reliable resource-monitoring to peers by using both oracle and measurements at nodes.

On the other hand, the strength of \emph{tit-for-tat} mechanisms is that every peer can easily evaluate the amount of data provided by its neighbors. This information is trusted and very accurate while it requires no overhead at all. Moreover, it has been shown in several deployed systems that \emph{tit-for-tat} mechanisms are efficient to enforce incentives, as they are able to discriminate peer resources, giving advantages to nodes contributing the more to the system.

As concern data-driven peer selection, it is known to provide optimal performance for specific scenarios~\cite{massoulie.twigg.ea:randomized}, but it generates a lot of overhead and suffers of strong performance degradation if the neighborhood is restricted. Moreover, this selection scheme is very sensitive to \emph{cheating} because it is based on information provided by neighbors. In fact, a peer can largely increase the probability of being selected by simply advertising emptier chunk collections than actually possessed.

\section{Recursive approximations}

We propose in this section to derive some recursive formulas that try to predict the behavior of the schemes that use the \emph{latest blind} chunk selection. This approach is similar to the one proposed in~\cite{sigepidemic,gyorgy:delay} in the case of homogeneous peers.

\subsection{Understanding chunk diffusion in heterogeneous networks}
\label{sec:hete_diff}

It has been shown in~\cite{sigepidemic} that agnostic diffusion schemes, which do not take into account peer resources when performing the peer selection, degrade their performance in heterogeneous upload capacity scenarios. One of the keys to produce accurate recursive formulas is to understand the reasons of this performance degradation and to identity the main issues for chunk dissemination in heterogeneous systems. 

To illustrate the performance degradation, we consider a simple system composed of $n=600$ peers and a source. We suppose $T_{SR}=T_S=1~s$ so that the source generates and uploads one chunk per second, and that peers have a buffer of 50 seconds.  We investigate  two scenarios: a first one, called homogeneous, where all peers have $u=u_S$; a second one, called heterogeneous, where $400$ peers have an upload capacity of $u_1=0.5~u_S$ ($T_1=2~s$) while the remaining have an upload capacity of $u_2=2~u_S$ ($T_2=0.5~s$). Note that the average bandwidth is the same in both scenarios. 
\begin{figure}[ht]
\begin{center}
\subfigure{\includegraphics[width=0.45\textwidth]{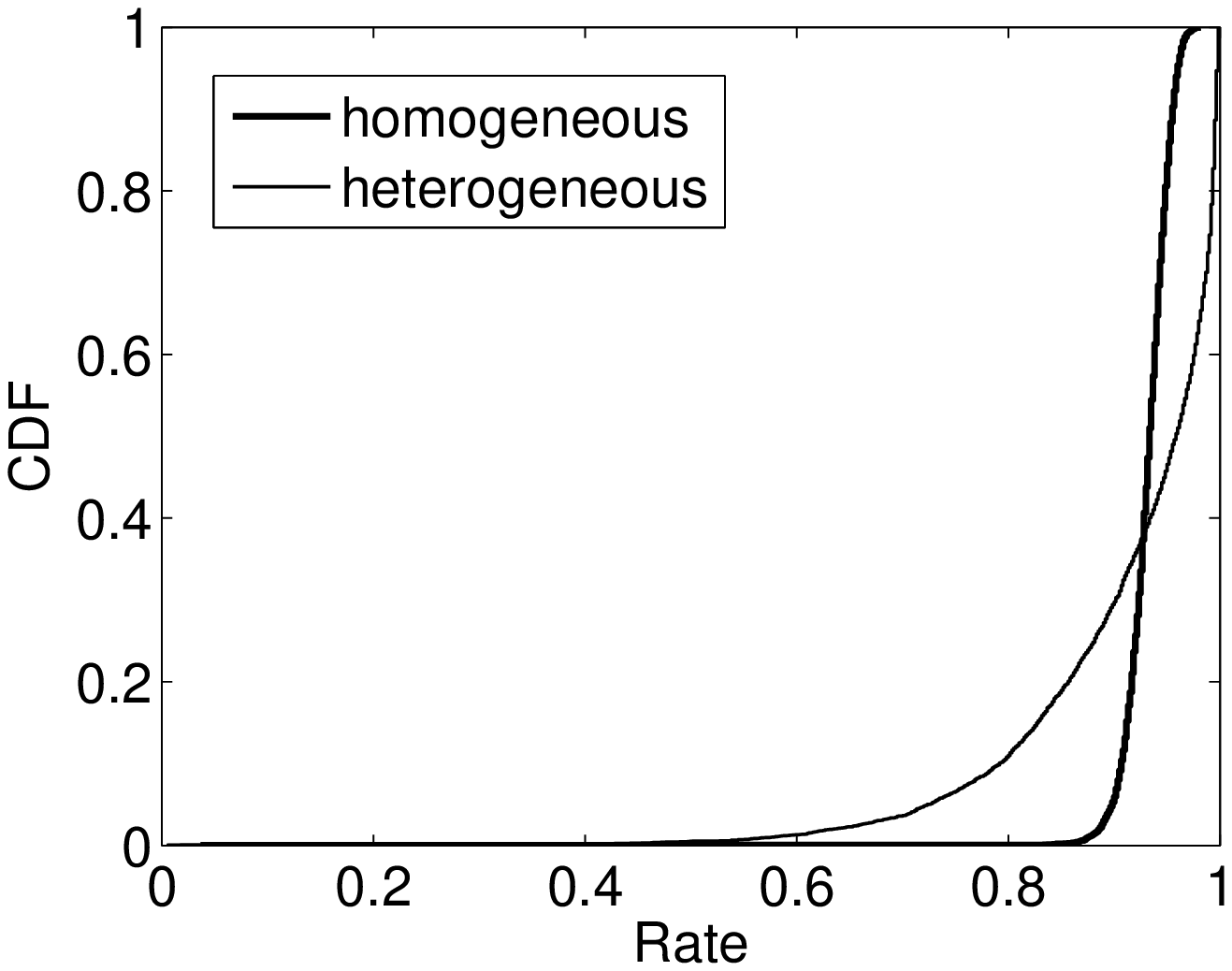}\label{fig:h_rate_comp}}
\subfigure{\includegraphics[width=0.45\textwidth]{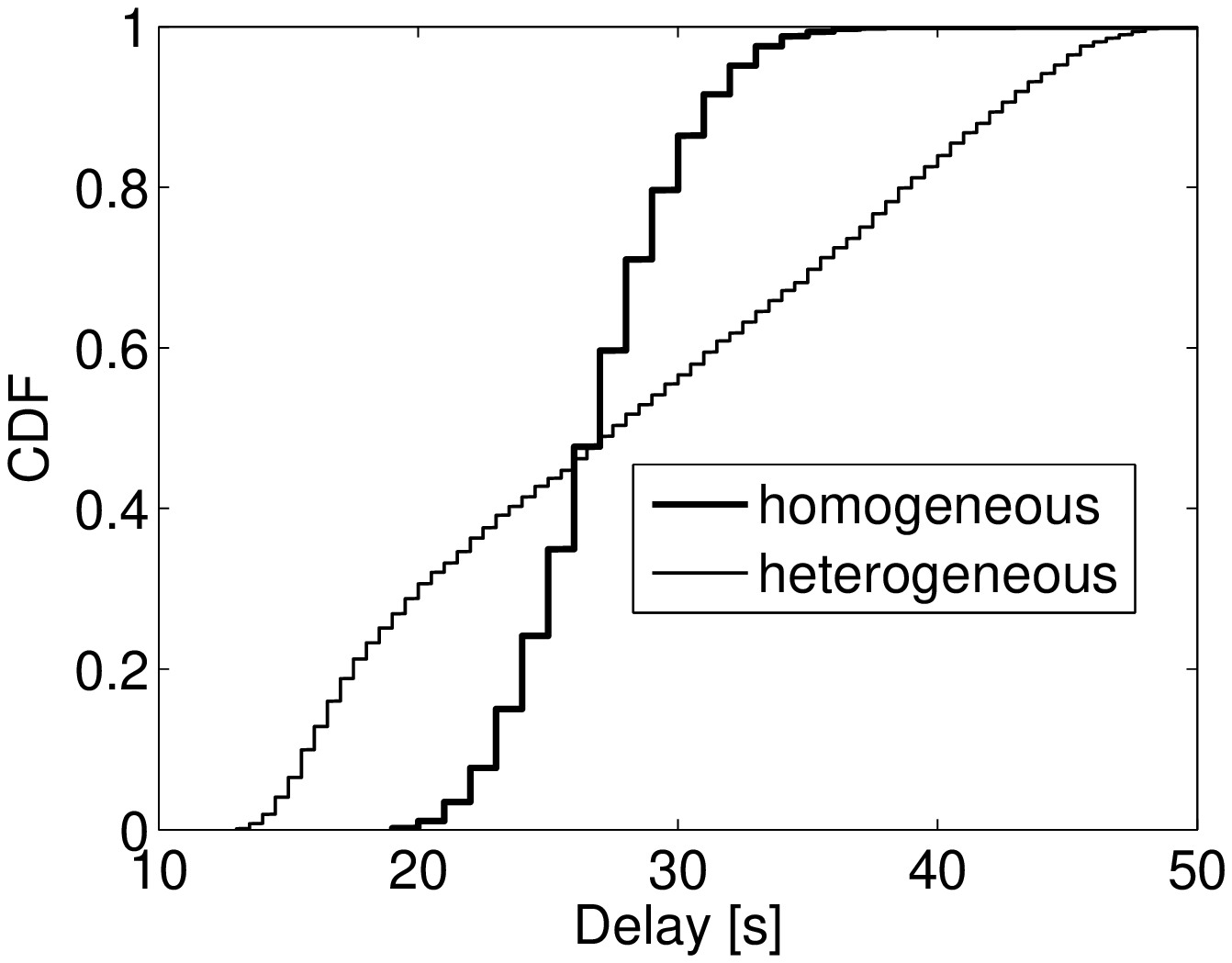}\label{fig:h_d95comp}}
\caption{CDF of chunk diffusion performance in case of homogeneous and heterogeneous upload capacities for the \emph{RP/LU} scheme.}
\label{fig:heterogeneity_CDF}
\end{center}
\end{figure}

In figure~\ref{fig:heterogeneity_CDF} we report the CDF of the chunks' diffusion rate/delay for the \emph{RP/LU} scheme. In the homogeneous case, the distributions are tightly concentrated around their averages ($25$ seconds for delay, and $0.93$ for the rate), while in the heterogeneous case, they are scattered over a larger range of values. 

\begin{figure}[ht]
\begin{center}
\subfigure{\includegraphics[width=0.45\textwidth]{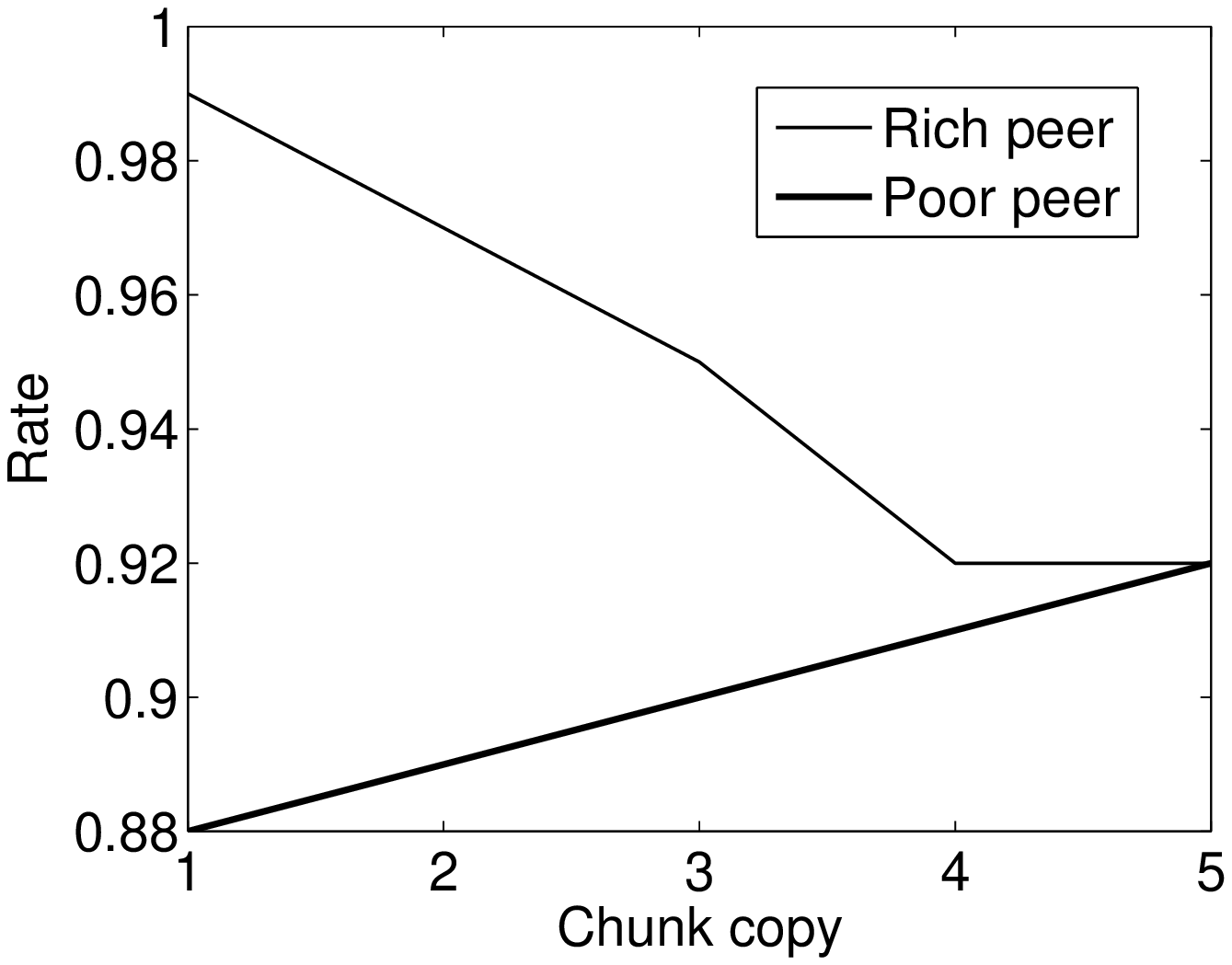}\label{fig:h_rate_comp2}}
\subfigure{\includegraphics[width=0.45\textwidth]{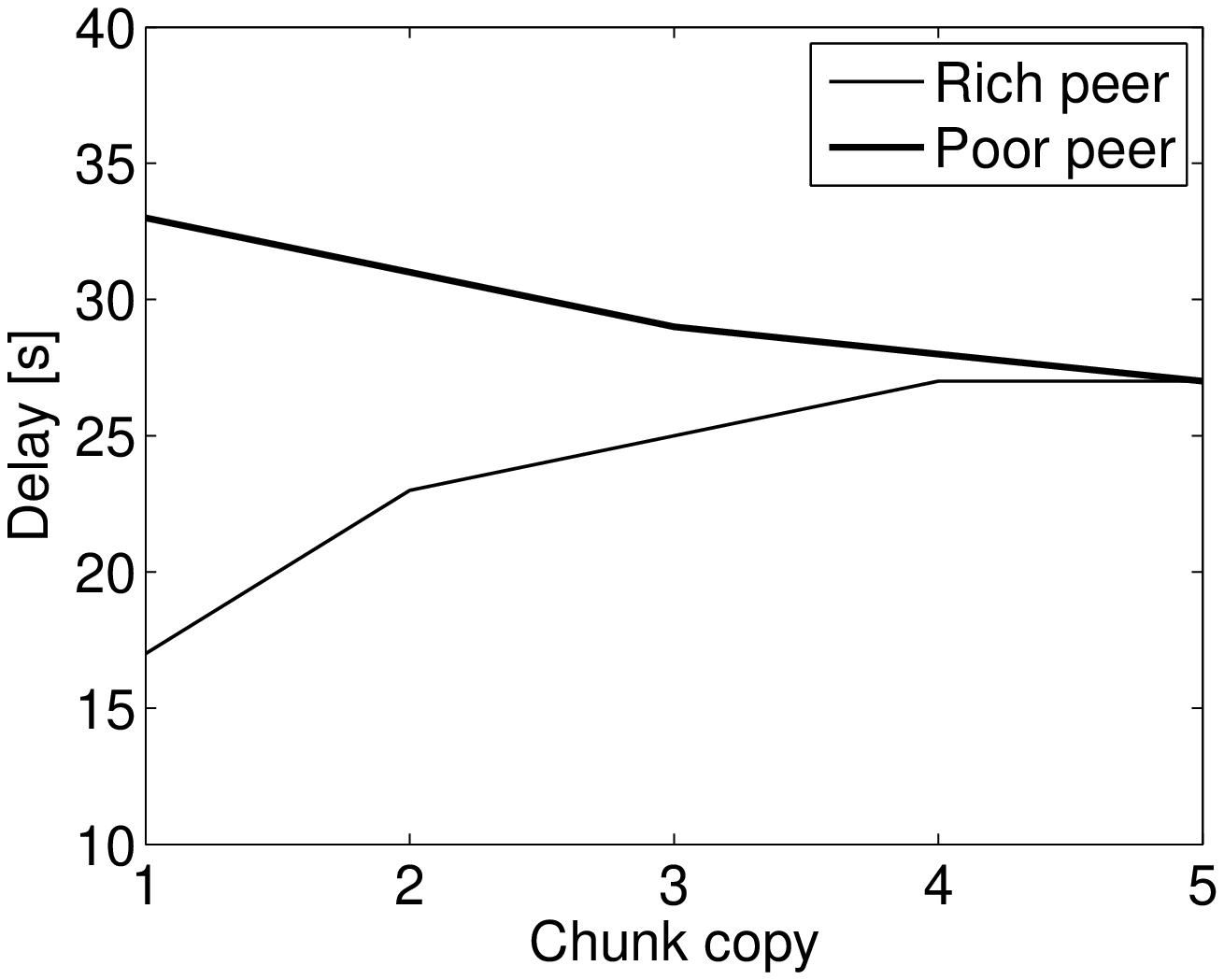}\label{fig:h_d95comp2}}\caption{Rate/delay performance for the \emph{RP/LU scheme} as a function of the resources of the $k^{th}$ peer receiving a given chunk. Rich peer $u(l)=2~u_S$, Poor peer $u(l)=0.5~u_S$.}
\label{fig:heterogeneity_copy}
\end{center}
\end{figure}

In order to better understand this behavior, we analyze the impact the resources of the first peers receiving a given chunk have on the final diffusion performance. For a given copy number $k$, Figure~\ref{fig:heterogeneity_copy} shows the rate/delay performance of a chunk depending on whether its $k^{th}$ copy has been received by a rich peer ($u(l)=2$; the thin curves) or by a poor peer ($u(l)=0.5$; the bold curves). We observe very different diffusion rate/delay performance, especially for the earlier copies. This difference lowers with the number of chunk replicas up to the $5^{th}$ copy, after which the resources of the receiver do not significantly affect the final rate/delay values. This clearly indicates that in the heterogeneous case, the diffusion performance is strongly impacted by the bandwidths of the first actually selected peers, while after a certain number of copies this impact is very limited. We claim that the scattered performance distribution in the heterogeneous case is mainly due the random selection of the first chunk exchanges, that leads to different performance according to the resources of the selected peers. 

All these results highlight the importance of resource awareness in peer selection. The intuition is confirmed: the first copies of a chunk should be exchanged between nodes with higher upload capacities in order to have diffusion trees with wide first levels. This reduces the diffusion delay and increases the number of peers that receive a given chunk before fresher ones are spread in the system. 

\subsection{Recursive formulas}\label{sec:epidemics_aware_formulas}

We now explain how to derive recursive formulas for a generic diffusion scheme based on an \emph{aware peer} selection coupled with a \emph{latest blind chunk} selection. The \emph{latest useful} selection, for which we do not provide formulas in this paper, will be the subject of the next section.
As we have just seen, in the case of heterogeneous upload capacities, the rate/delay distribution is not centered around a given value but scattered over a large range. In order to approximate the diffusion functions in such scenarios, it is therefore more significant to work with distribution instead of using only averaged values (which suffices in the homogeneous case~\cite{sigepidemic,gyorgy:delay}).

As the performance is mainly affected by the first chunk exchanges, we propose a two-step approach: first an exact description of the early behavior of the diffusion, then the use of averaged approximation to derive the rest of the diffusion process.

Let $J$ be a distribution of system states that describes the early behavior of a chunk's diffusion. One may think of $J$ as the \emph{initial conditions} of the diffusion. These initial conditions represent different possible evolutions of first chunk exchanges up to a certain time $T_{init}$. We propose to use $J$ to compute a recursive approximation of the afterwards diffusion. The larger (and the more accurate) the initial conditions are, the better the distribution computed by the recursive formulas will fit the real distribution.

The initial conditions should be deterministically computed according to the diffusion scheme (see below); such operation can be computationally expensive and exponentially time consuming (we have to limit ourselves to the early diffusion). However, as we observed, most of the variance in the diffusion process is captured by the very few first exchanges; this keeps the approach proposed here much less expensive in term of computational resources  and time than a complete simulative analysis.

We assume a scenario where every peer has a complete knowledge of the overlay (full mesh connectivity) and that the $H$ and $W$ parameters are the same for all peers. We also suppose that the resources shared by a node are defined by its class, so we can express the probability that a peer of class $i$ selects a peer of class $i'$ as $\beta(i,i')$.

As for the recursive formulas derived in~\cite{sigepidemic} we assume that the number of peer is sufficiently large, so that the system may be considered in the mean field regime where peers are mutually independent, and that the probability that a given chunk belongs to $B(l)$ is independent from the fact that any other chunk belongs to $B(l)$ (the validity of these assumptions will be checked later).

We make the approximation that all peers of the same class are synchronized in uploading a chunk. $0$ being the time of one given chunk's creation, we define $\T_{i}:=\left \{T_{i}~2T_{i}~3T_{i}~..\right \}$ as the set of times at which peers of class $i$ may send a chunk, and $\T_{SR}:=\left \{T_{SR}~2T_{SR}~3T_{SR}~..\right \}$ as the set of chunk generation times. We define $\T=\T_1 \cup \T_2 \cup ... \cup \T_{U} \cup \T_{SR}$ as the (sorted) set of times at which an event occurs. Simultaneous events from distinct classes are taken into account with their multiplicity.

The values we are interested in are the fraction of the peers of a class that received the chunk before time $t$. For every instant of time $t \in \T$ and each class $i$, we propose to compute that fraction, denoted as $r_i(t)$.

  
The first step is to compute the initial conditions $J$. A set of $|J|$ instances of the $r_i(T_{init})$ are generated according to the considered scheme. Note that for an instance $j\in J$, all $r_i(T_{init})$ are deterministic.
Starting from these initial conditions the recursive formulas describe the diffusion function for each $j\in J$. In the following when considering a given $r_i(t)$, we assume implicitly an initial condition $j\in J$, while the average over $J$ is denoted as ${\overline r_i(t)}$.



For every time $t \in \T : t > T_{init}$ at which an upload event occurs, we denote as $i$ the class sending the chunk at that time $t$, and as $t'$ the instant of time preceding $t$ in $\T$. 
We denote as $p(t)$ the probability that a given peer ends the upload of the chunk at time $t$, so that on average $np(t)$ transmissions of the considered chunk finish at time $t$. 
$p(t)$ is initially set to $0$ for all $t$ values.  
That probability $p(t)$ is spread over the U classes according to the selection probability $\beta$, so that peers in class $k$ receive the tagged chunk at time $t$ with probability $\beta(i,k)p(t)$. Among a given class target peers are then selected uniformly at random. Due to this random selection, the number of copies of the tagged chunk that are received by an arbitrary peer is a binomial random variable with parameter ($\alpha_kn,\beta(i,k)p(t)/\alpha_kn$). For large $n$, this can be approximated by  a Poisson random variable with mean $\beta(i,k)p(t)$. The probability that a peer of class $k$ receives at least one copy of the tagged chunk at time $t$ is therefore approximately equals to $1-e^{-\beta(i,k)p(t)}$. A fraction $1-r_i(t)$ of the peers that receive the chunk at time $t$ actually need it. The recursive formula is then:

\begin{equation}\label{eq:awlb}
\forall k : 1 \le k \le U,
r_k(t)=r_k(t')+(1-e^{-\beta(i,k)p(t)})(1-r_k(t'))
\end{equation}

We then need to update the value of $p(t)$ for the later event in $\T_i$. This means to compute the probability that the chunk is the latest in the collection of chunks $B$ of peers of class $i$. This affects the probability that the download of the tagged chunk ends at time $t+T_i$ as follow:

\begin{equation}	\label{eq:bt_hete} 
p(t+T_i)=p(t+T_i) + \alpha_i r_i(t)\prod^{\lfloor\frac{t}{T_{SR}}\rfloor}_{k=1}(1-{\overline r_i(kT_{SR}}))
\end{equation}

For every time $t \in \T_{SR} : t > T_{init}$, at which a new chunk is generated, the status of the considered chunk is unchanged (no transmissions occur for it) so we simply have:

\begin{equation}\label{eq:awlb_ts}
\forall k : 1 \le k \le U\text{,}
r_k(t)=r_k(t')
\end{equation}

\subsection{Formulas validation}
We validate the recursive formulas by considering the BA peer selection process with awareness probability $W=1$. We suppose the overlay is a complete graph and the source injects only one copy of each chunk in the system ($T_{SR}=T_S$). To this goal we set the chunk size to $c=0.9~Mb$ and the source upload capacity to $u_S=0.9~Mbps$. The other parameters are those of the reference scenario described in the next section.

We consider two different sets of initial conditions: $J_1$ and $J_2$. The former is composed of only one initial condition ($|J_1|=1$), and it is only based on the copy uploaded by the source ($T_{init}=T_{SR}$). In this case, we will only have one rate/delay value and not a distribution. The latter is composed of $|J_2|=1000$ different initial conditions, and is based on $T_{init}=T_{SR}+1~s$ (given the system parameters used, an initial condition represents 5 chunk exchanges on average). In this case, we will have a distribution based on 1000 different chunk diffusions.

\begin{figure}[t]
\begin{center}
	\subfigure{\includegraphics[width=0.45\textwidth]{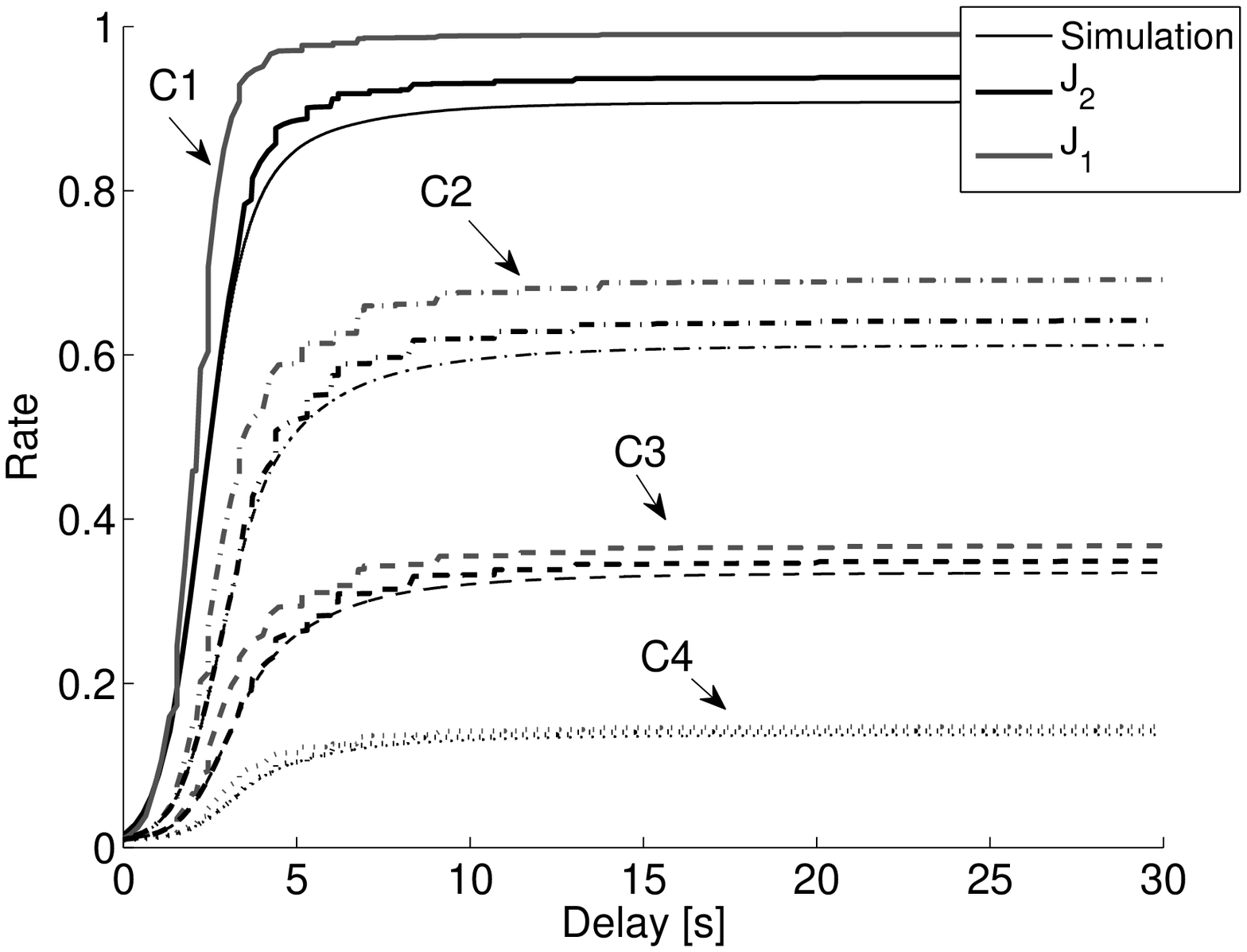}}
\subfigure{\includegraphics[width=0.45\textwidth]{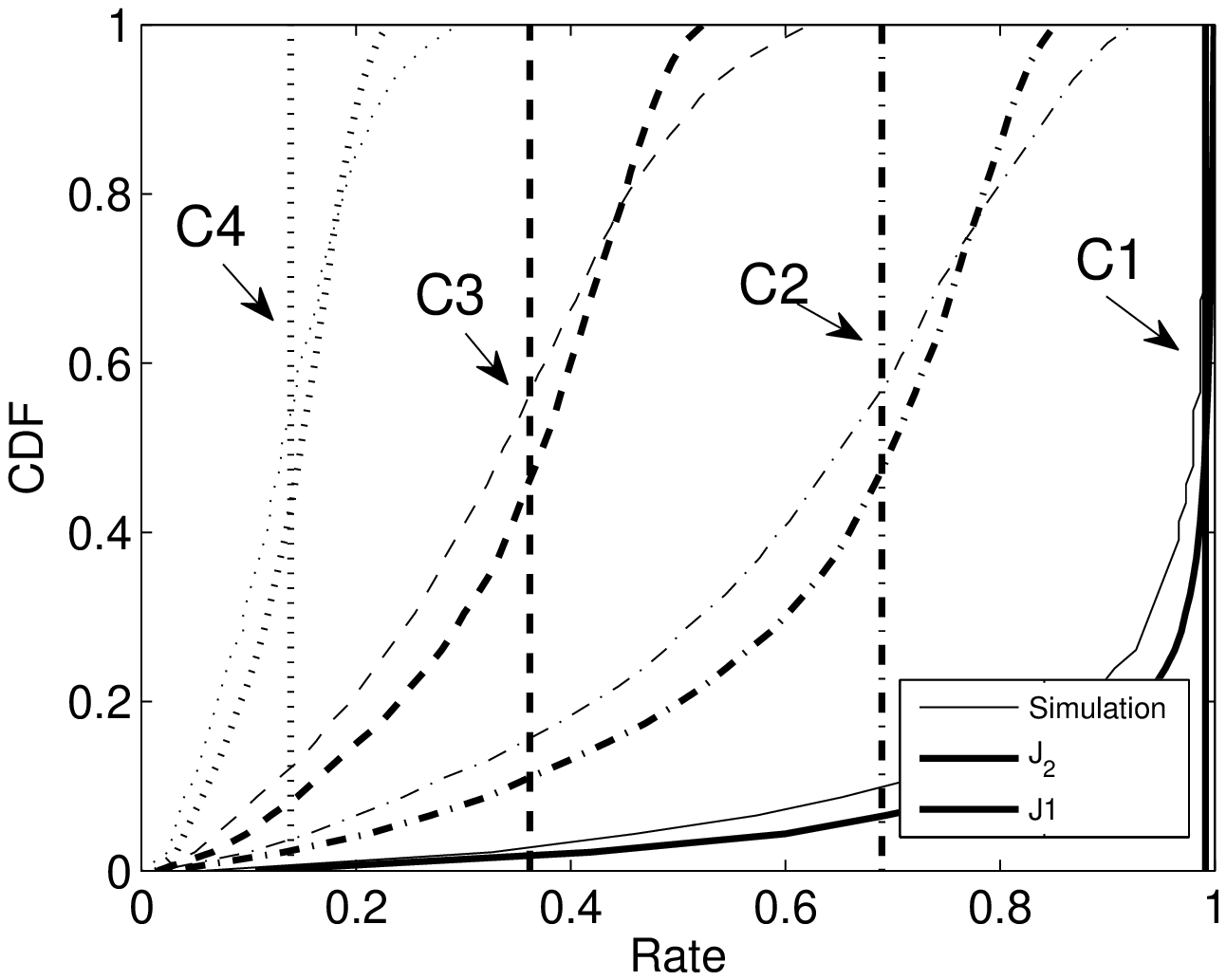}}
	\caption{Per class validation of the recursive formulas. BA peer selection.}\label{fig:epidemics_aware_validation}
\end{center}
\end{figure}

Figure~\ref{fig:epidemics_aware_validation} shows formulas are quite accurate in predicting the rate/delay performance of the considered scheme. As expected, to increase the number of initial conditions and $T_{init}$, increases the accuracy of the performance prediction. In particular, the distribution based on $1000$ samples of 5 chunk exchanges fits pretty well the distribution based on a simulation of 10000 chunks. It is possible to observe estimation errors between 0-7\% (C4-C2) as concern diffusion rate, and 10-15\% (C1-C4) as concern the average delay. 

These errors are slightly larger than in the homogeneous case studied in~\cite{sigepidemic}. This is due to the variability of the diffusion process that is more stressed in heterogeneous systems because of the additional randomness of the different upload capacities. Nevertheless the obtained results are worthwhile for having a fast performance estimate of a system.

\section{Performance analysis}
\label{sec:simus}

In this section, we evaluate the rate (or miss ratio)/delay trade-off achieved by resource aware selection schemes.  In particular, we focus on the performance of three representative peer selection policies:  \emph{random peer} (RP), \emph{bandwidth-aware} (BA) and \emph{tit-for-tat} (TFT).

To this purpose we use an event-based simulator developed by the Telecommunication Networks Group of Politecnico di Torino\footnote{\url{http://www.napa-wine.eu/cgi-bin/twiki/view/Public/P2PTVSim}} where we implement the aforementioned schemes.

Unless otherwise stated, we suppose there are $n=1000$ peers and we set their uplink capacities according to the distribution reported in Table~\ref{tab:epidemics_bw_distrib}, that is derived from the measurement study presented in~\cite{distrib}, and that has been used for the analysis in~\cite{guo.y:adaptive}. We suppose $p_e=0.05$ so that every peer has about 50 neighbors, $N(l)\approx50$. The source has about 50 neighbors as well, an upload capacity $u_S=1.1~Mbps$ and employes a RP selection policy. 

In order to avoid critical regime effects, we suppose the stream rate $SR=0.9~Mbps$ that leads to a bandwidth balance of $1.13~SR$.  We set the chunk size $c=0.09~Mb$, we suppose peers have a buffer of 30 seconds and for the TFT scheme the epoch length is set to $T_e=10~s$.  

The chunk selection policy we consider here is \emph{latest useful}.
\begin{table}[ht]
	\centering
\
\begin{tabular}{|c|c|c|}
		\hline
		Class & Uplink [Mbps] & Percentage of peers\\
		\hline
		C1& 4 & 15\%\\
		\hline
			C2 & 1 & 25\%\\
		\hline	
		C3 & 0.384 & 40\% \\
		\hline
		C4 & 0.128 & 20\%\\
		\hline			
		\end{tabular}
		\caption{\label{tab:epidemics_bw_distrib}Upload capacity distribution with mean 1.02 Mbps.}\vspace{-0.8cm}
\end{table}

\subsection{Reference scenario}
\begin{figure}[t] 
\begin{center}
\subfigure[Class $C1$]{
	\includegraphics[width=0.4\textwidth]{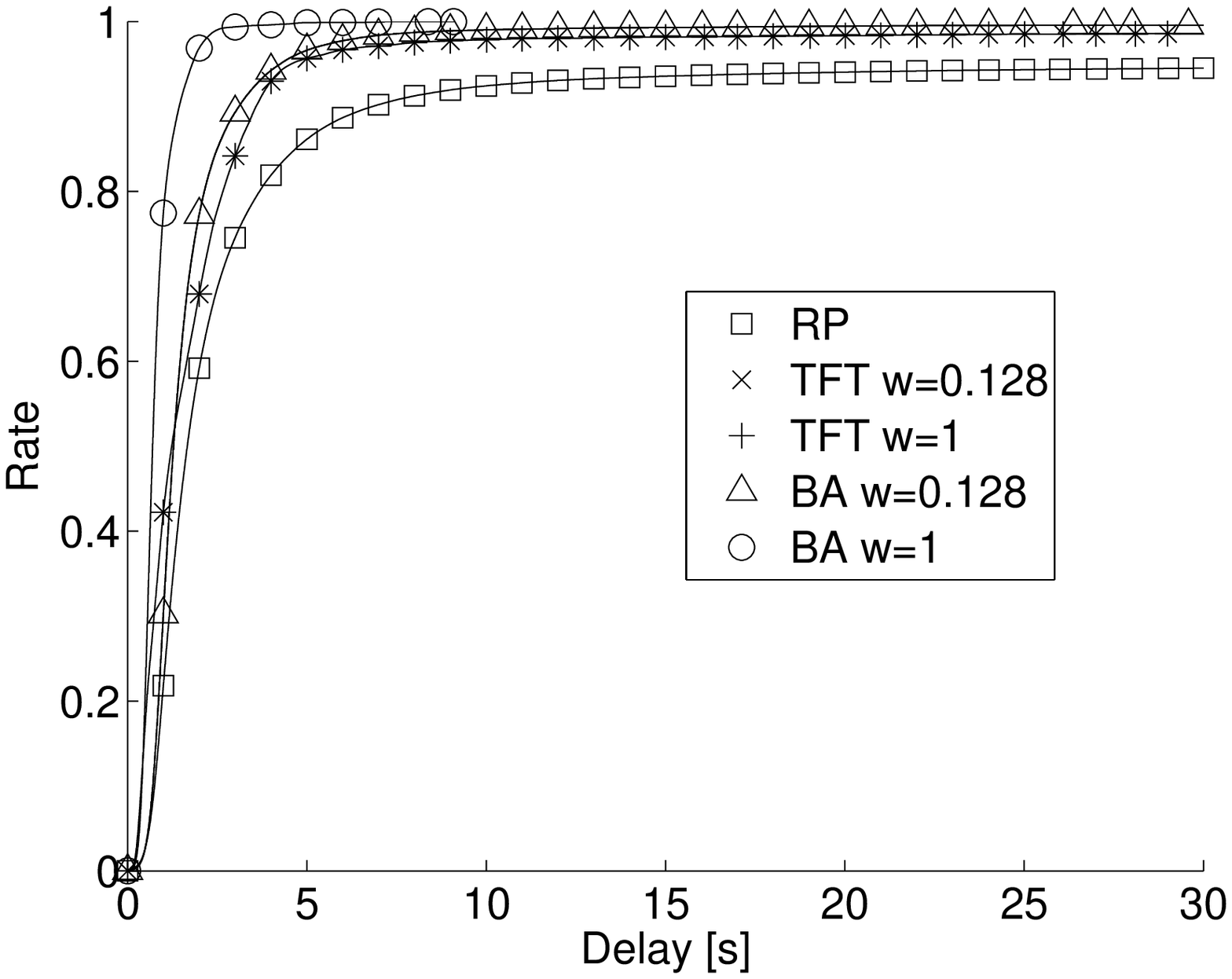}
}
\subfigure[Class $C2$]{
	\includegraphics[width=0.4\textwidth]{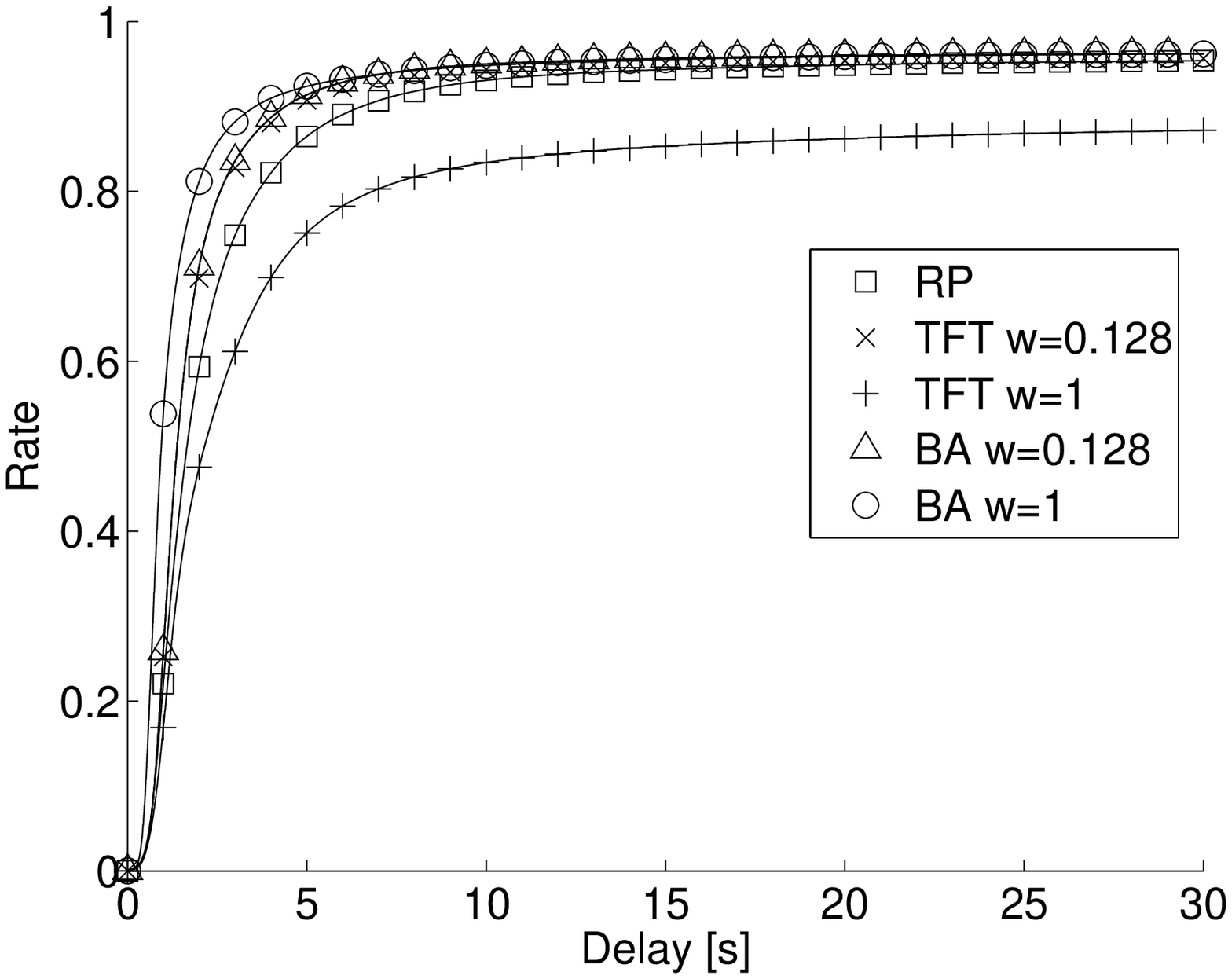}
	}\\
\subfigure[Class $C3$]{
        \includegraphics[width=0.4\textwidth]{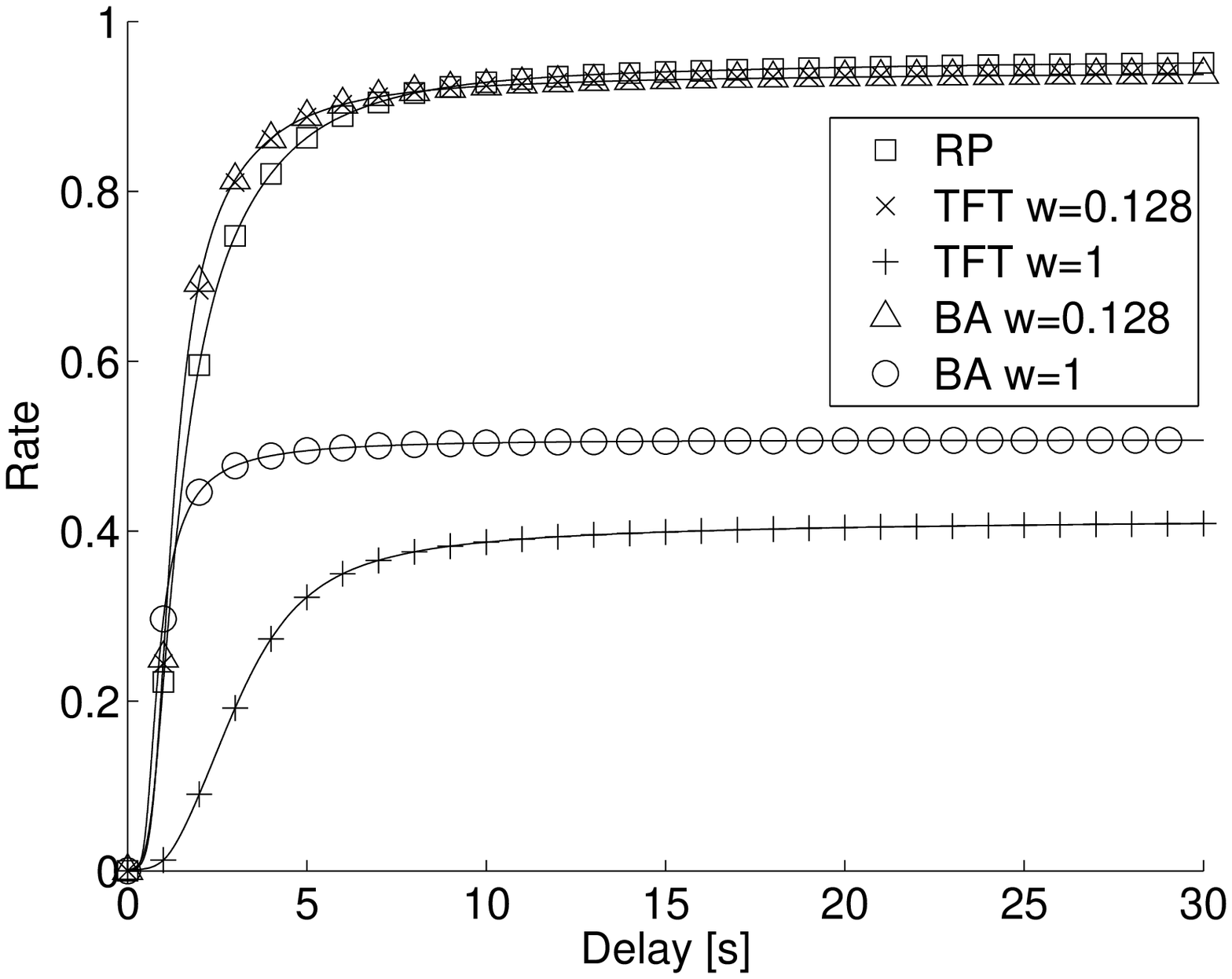}
}
\subfigure[Class $C4$]{
        \includegraphics[width=0.4\textwidth]{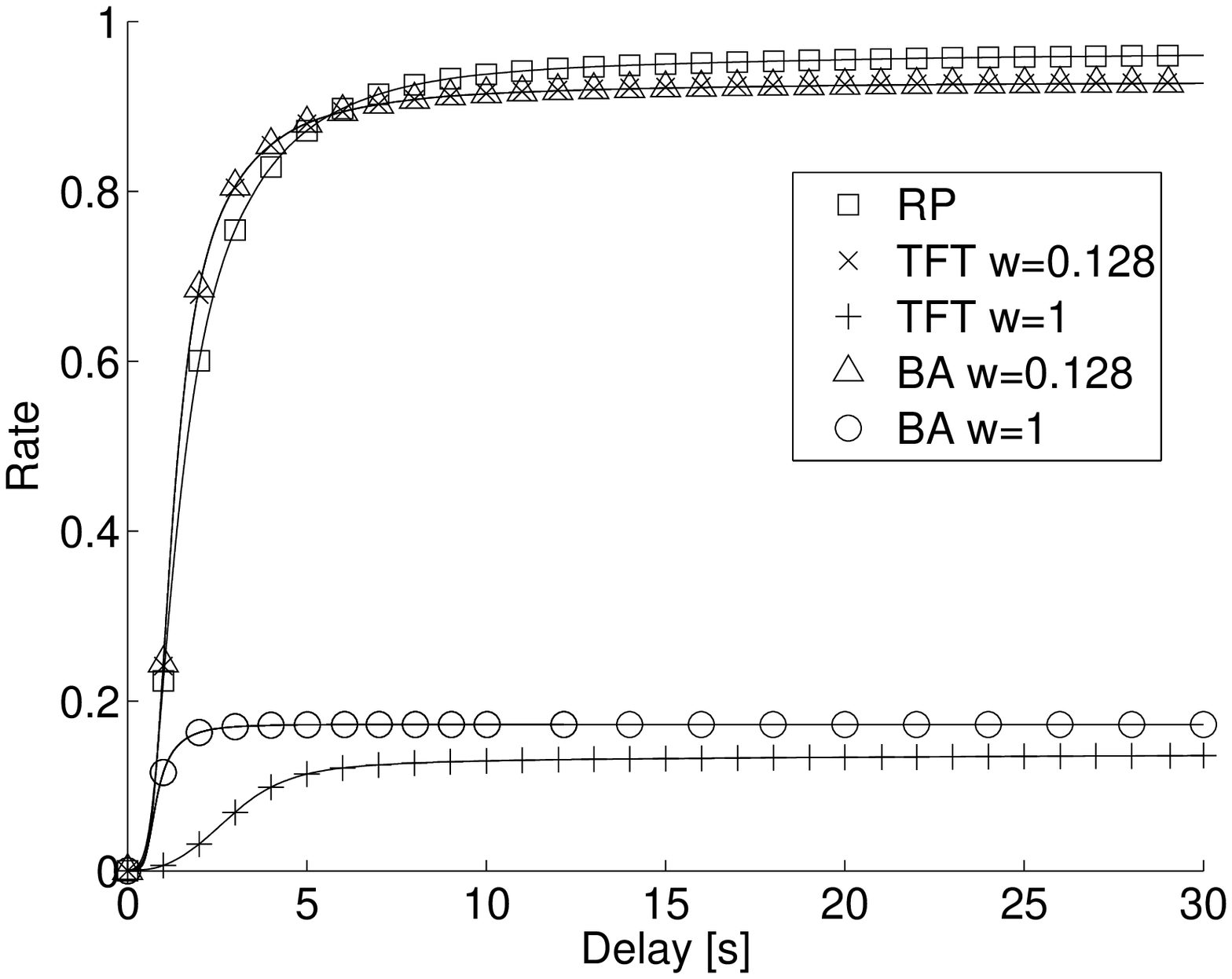}
}
\caption{Chunk diffusion in the reference scenario}\label{fig:epidemics_aware_reference}
\end{center}
 \end{figure}
 
We first consider a reference scenario whose diffusion process of the different schemes is pictorially represented in Figure~\ref{fig:epidemics_aware_reference} for all classes. For BA and TFT peer selection we consider two values of awareness probability: $W=1$ and $W=0.128$ corresponding to a fully-aware and a generous approach respectively.
 
We observe schemes taking into account peer contributions/resources in general decrease the diffusion delay with respect to the agnostic RP for all classes. BA gives priority to richer peers, so that the diffusion process is speeded up thanks to their high upload capacity placed at the top of chunk diffusion trees. On the other hand,  TFT clusters peer according to their resources~\cite{gai.a.mathieu.f.ea:stratification}, leading to a similar effect as the one observed in the experimental analysis of incentive-based live streaming systems~\cite{pulseawareness}.

Such resource aware schemes increase the diffusion rate of the richer classes C1-C2, while they reduce the one of poorer classes C3-C4. This rate decrease is particularly dramatic in case of a completely aware selection (W=1). On the other hand, if the selection is more generous (W=0.128), this drastic reduction is avoided, but the diffusion delay may increase, especially if the BA selection is used.

This clearly highlights a rate/delay trade-off as a function of the awareness probability $W$.

\subsection{Awareness-Agnostic peer selection trade-off}
\label{sec:epidemics_aware_trade_off}
\begin{figure}[ht] 
\begin{center}
\subfigure{\includegraphics[width=0.45\textwidth]{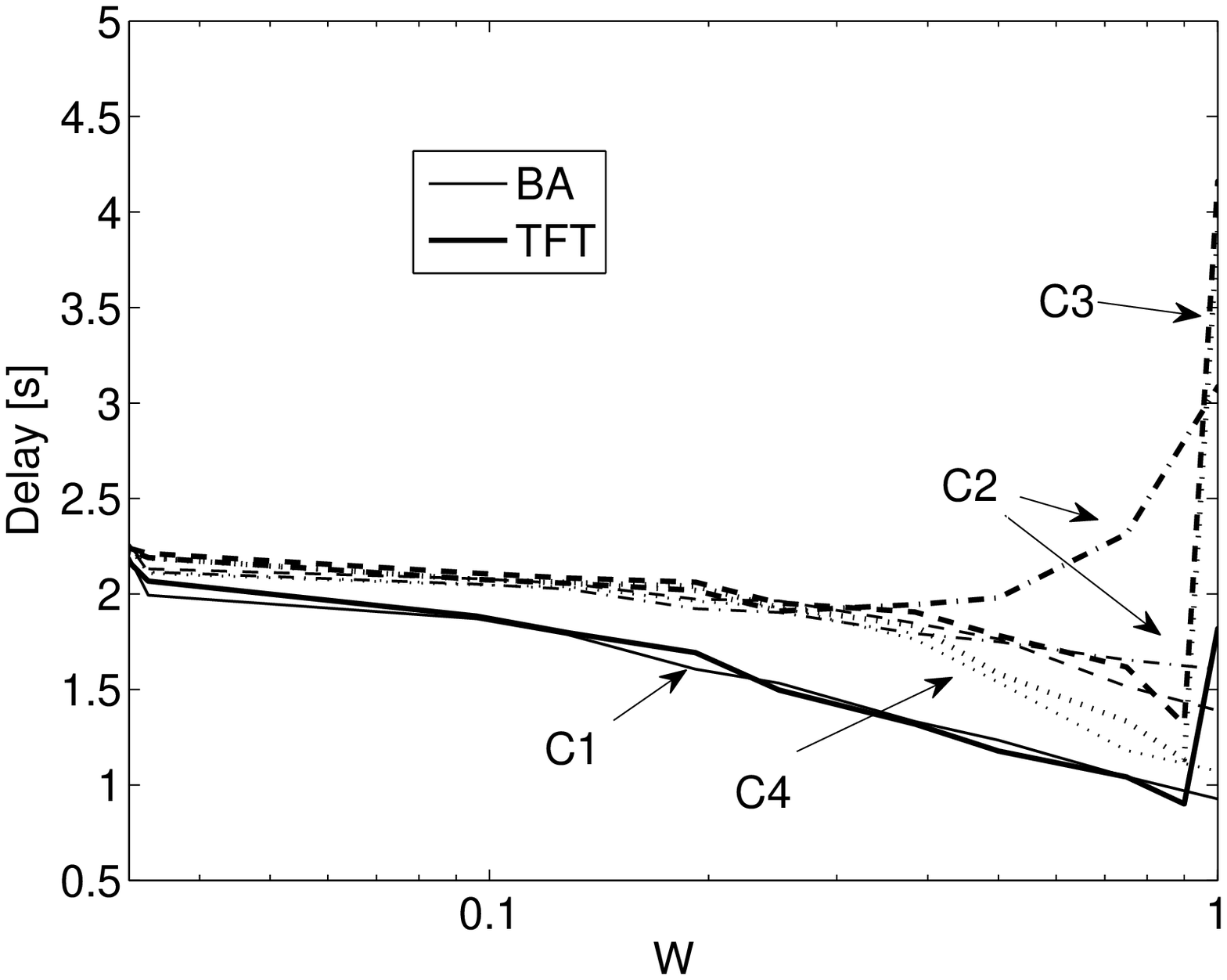}}
\subfigure{\includegraphics[width=0.45\textwidth]{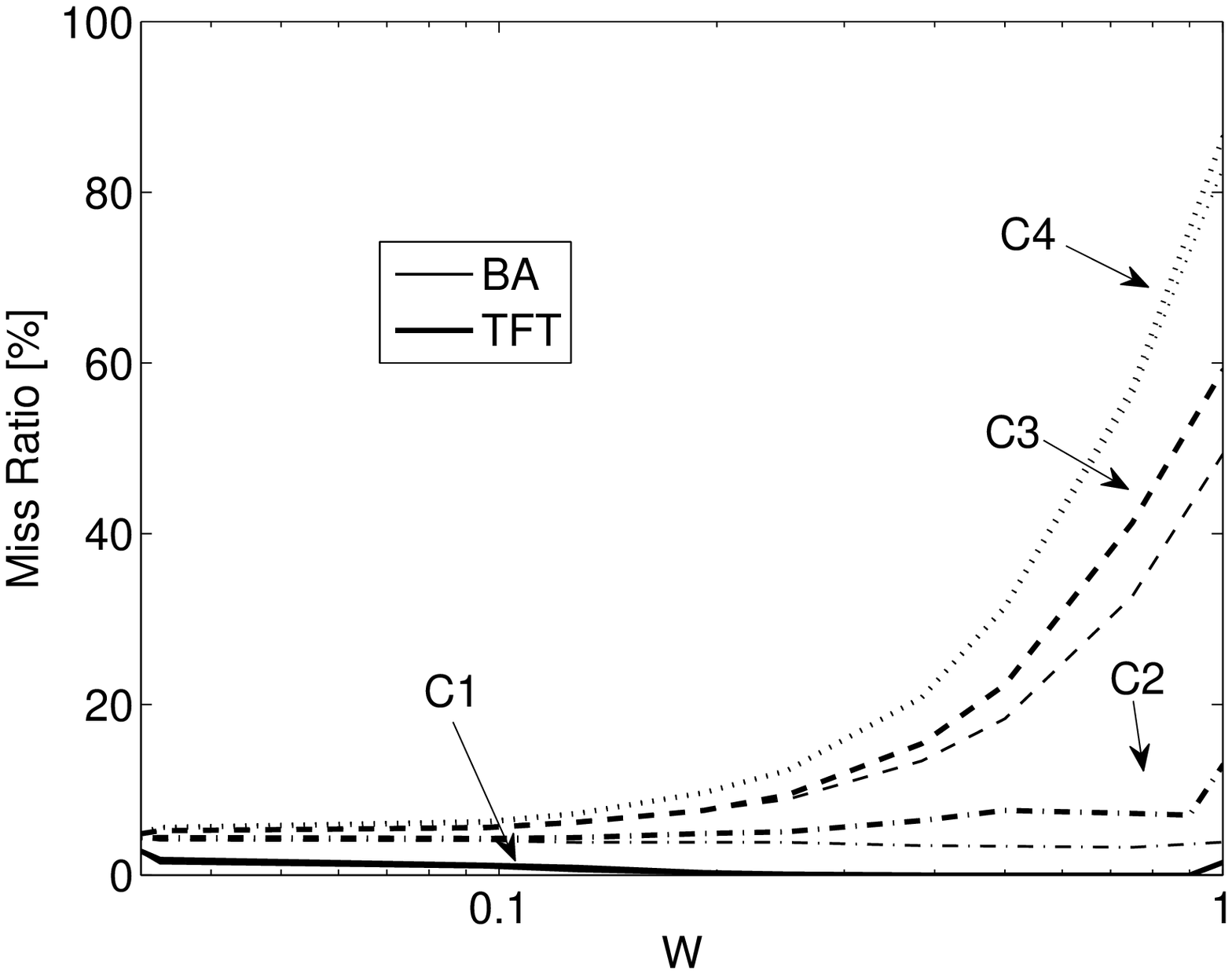}}\vspace{-0.4cm}
\caption{Diffusion delay and chunk miss ratio as a function of the awareness probability.}\vspace{-0.4cm}
\label{fig:epidemics_aware_tradeoff}
\end{center}
\end{figure}

Figure~\ref{fig:epidemics_aware_tradeoff} reports the rate/delay performance of $BA$ and 
$TFT$ schemes as a function of the awareness probability in the heterogeneous scenario described in Table~\ref{tab:epidemics_bw_distrib}. 

The diffusion delay decreases as the awareness probability increases for all bandwidth classes. 
This indicates the placement of the nodes with higher upload capacities at the top of the diffusion trees effectively speeds up the diffusion process.
We also notice that, by increasing the awareness probability, the delay differentiation between different classes increases as well. 
In particular, when $W\approx0$, all classes achieve the same diffusion delay because the selection is almost random (as in $RP$). On the other hand, when $W=1$ there is the maximum discrimination because the selection is purely aware.
In fact, more and more peers with higher upload capacities are selected first as the awareness probability increases.

Regarding the miss ratio, richer classes take advantage of the increasing awareness. On the other hand, the miss ratio of the poorer classes stagnates until a certain awareness value of about $W=0.22$, after which peers start missing more and more chunks. The intuition is that richer peers are selected with increasing frequency (decreasing their miss ratio), and the reverse for the poorer classes.

We observe that $BA$ scheme slightly outperforms $TFT$. This is not surprising: BA weights peers according to their upload capacity, so that it perfectly discriminates them according to their resources. However, the gap is very small making $TFT$ appealing  for real deployment  because more simple and reliable than $BA$.

Notice that a pure $TFT$ approach ($W=1$) performs poorly: without agnostic disseminations, the peer clustering generated by $TFT$ interferes with a proper dissemination of the chunk among all the peers of the system. This does not happen under $BA$ scheme because every peer can be selected with low probability, even poorer ones, assuring that every chunk can eventually reach all peers.

 \begin{table}[ht]
	\centering
\
\begin{tabular}{|c|c|c|}
		\hline
		Class & Uplink [Mbps] & Percentage of peers\\
		\hline
		$\tilde{C1}$& 3.5 & 7\%\\
		\hline
		$\tilde{C2}$ & 0.35 & 66\%\\
		\hline	
		$\tilde{C3}$ & 0.2 & 27\% \\
		\hline			
		\end{tabular}
		\caption{\label{tab:epidemics_bw_distrib_turin}Upload capacity distribution with mean 0.53 Mbps.}\vspace{-0.4cm}
\end{table}


In order to validate our claims, we consider another bandwidth distribution (Table~\ref{tab:epidemics_bw_distrib_turin}) which is derived from the measurement study presented in~\cite{ciullo.d:understanding}, and has been used for the evaluation of the BA principle  in~\cite{silva.leonardi.ea:bandwidth-aware}. We also consider the case of free-riders by setting the upload capacity of peers of class $\tilde{C3}$ to $0~Mbps$ instead of $0.2~Mbps$. In order to keep the same bandwidth balance as in the previous scenario, we reduce the stream rate to $SR=0.5~Mbps$, the chunk size to $c=0.05~Mb$ and the source upload capacity to $u_S=0.6~Mbps$. Note that in this scenario the bandwidth distribution is more skewed. 
Since the two selection policies behave similarly, in the following we focus on $TFT$ peer selection.

\begin{figure}[ht]
\begin{center}
\subfigure{\includegraphics[width=0.45\textwidth]{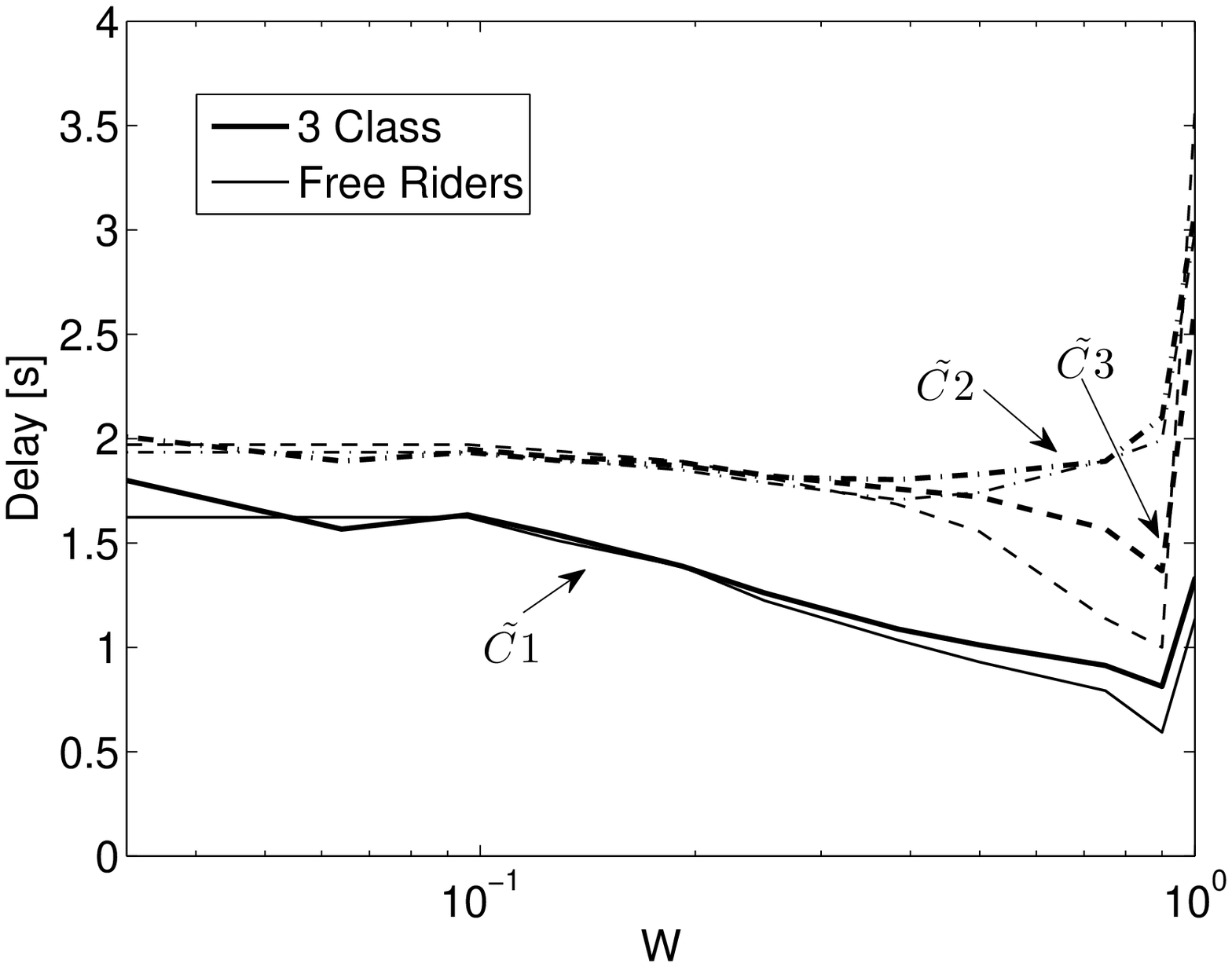}}
\subfigure{\includegraphics[width=0.45\textwidth]{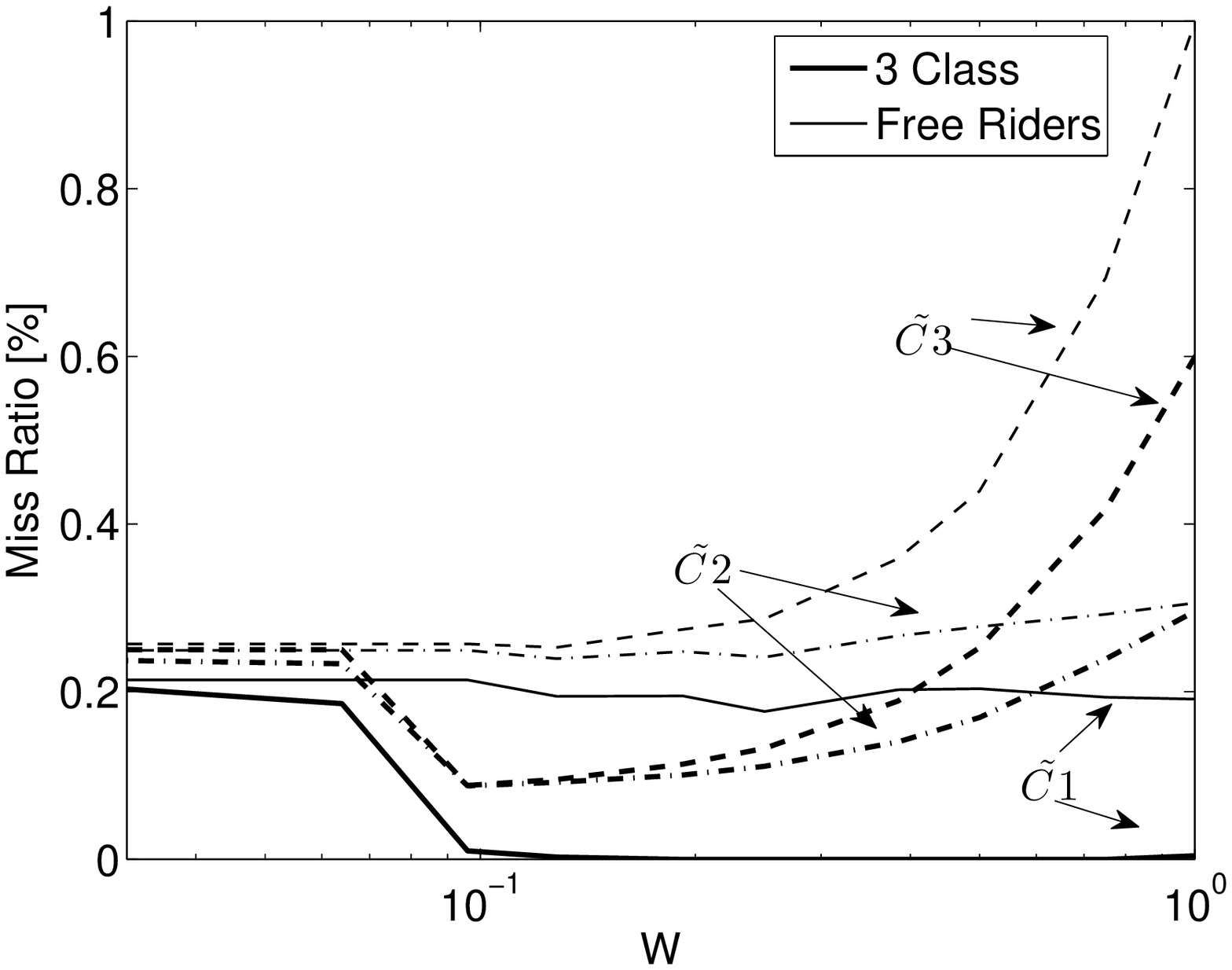}}\vspace{-0.3cm}
\caption{\emph{TFT} performance as a function of awareness parameter for a skewed bandwidth distribution and in presence of free-riders.}\vspace{-0.2cm}
\label{fig:epidemics_aware_h}
\end{center}
\end{figure}

Figure~\ref{fig:epidemics_aware_h} highlights the trend in the \emph{3 classes} scenario is similar to the one observed before. The only difference is that the gain of the increasing awareness is more evident for all classes. This is due to the high bandwidth of the first class with respect to the stream rate: as soon as this class is privileged all peers improve their performance.

In the scenario with \emph{free-riders}, all chunks the source uploads to class $\tilde{C3}$ are lost because peers cannot upload them.  So the miss ratio cannot be lower than the percentage of peers of class $\tilde{C3}$.  Classes $\tilde{C1}$ and $\tilde{C2}$ almost receive all the other chunks while free-riders are identified and receive a decreasing percentage of data as the awareness probability increases. 
This highlights that, in an heterogeneous scenario, the selection policy employed by the source can have a tremendous impact on the system performance. If the source could discriminate peers according to their resources, we won't observe such a miss ratio. We better investigate in the following the impact of different source selection schemes.

In all scenarios we observe the presence of a minimum suitable value of awareness probability. In fact, it is not interesting to select an awareness probability $W < 0.1$ because there is almost no gain with respect to the $RP$ selection. From this value to $W=1$ ($W=1-\epsilon$ for $TFT$ scheme) a trade-off arises. The more the scheme is aware the more richer peers  improve their performance. On the other hand, even if there is enough bandwidth, peers of the poorer classes loose lot of chunks. 
This can be seen as a good property of the system because it incentives peers to contribute more in order to improve their performance. On the other hand, part of the bandwidth is lost. The best value for the awareness probability depends on the application environment but in any case this value should be larger than $0.1$ in order to discriminate peers according to their resources, to improve system performance and to recompense peers contributing the more.  

\subsection{Source scheduling}
We now analyze the impact of the source selection policy and of the source upload capacity on the scheme diffusion performance. 

\begin{figure}[t]
\begin{center}
\subfigure{\includegraphics[width=0.45\textwidth]{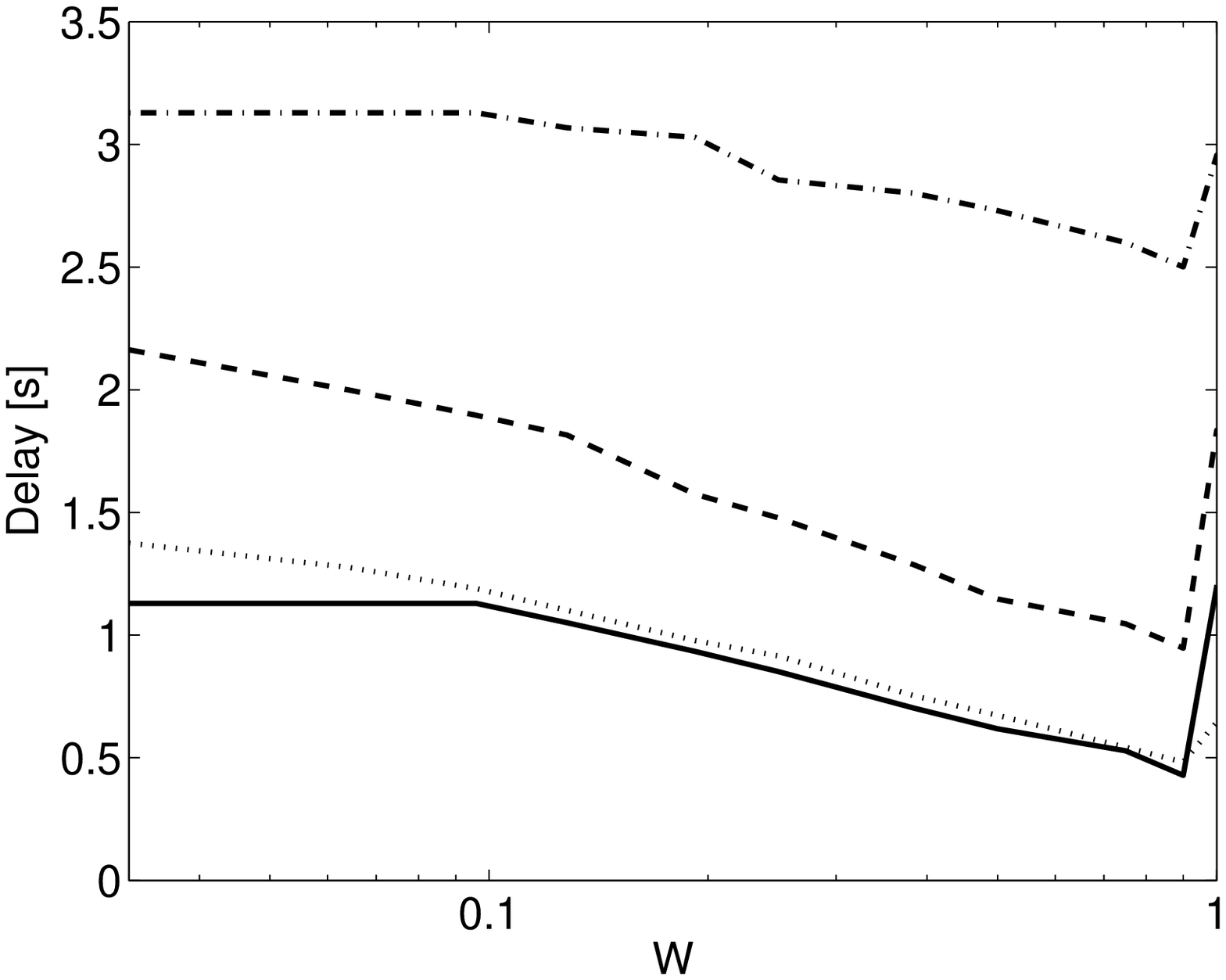}}
\subfigure{\includegraphics[width=0.45\textwidth]{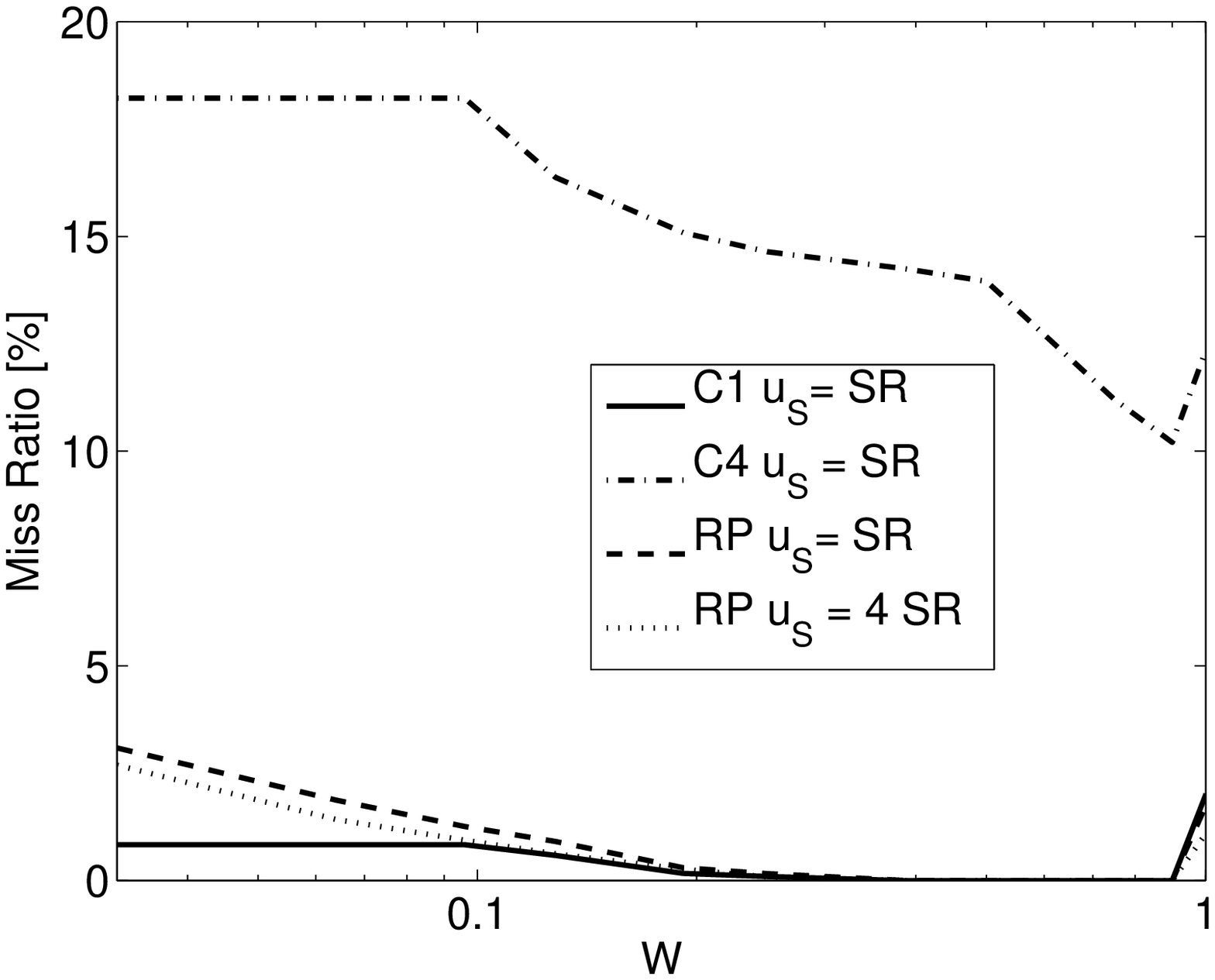}}\vspace{-0.3cm}
\caption{Diffusion delay and miss ratio of $C1$ peers as a function of awareness probability for different source selection polities. TFT selection at nodes.}\vspace{-0.5cm}
\label{fig:epidemic_aware_S1}
\end{center}
\end{figure}

In Figure~\ref{fig:epidemic_aware_S1}, we consider four different source policies: random peer selection ($RP$) with source upload capacity $u_S=SR$;  random peer selection with source upload capacity $u_S=4~SR$; selection of a peer of class $C1$ with upload capacity $u_S=SR$; selection of a peer of class $C4$ with upload capacity $u_S=SR$. We consider $TFT$ peer selection at nodes and, since the trend of all classes is similar, we only report in figure the performance of peers of class $C1$.

The diffusion delay strongly depends on the source policy. In fact, the selection of a peer of class $C1$ can reduce of 3 times the delay with respect to the selection of  a peer of class $C4$ while the $RP$ selection stays in between. But as explained earlier, it is very difficult to estimate the upload capacity of peers, and the source cannot employ a $TFT$ mechanism because it does not download any data. However, if the source has an upload capacity of $u_s = 4~SR$, the agnostic RP selection performs as the selection of a peer of class $C1$. This means that, if the source is slightly over-provisioned (remember that an upload capacity of $4~SR$ is negligible with respect to the number of peers), it has not to discriminate peers according to their resources. 

As for the concern miss ratio, we observe a dramatic degradation if the source sends the first copy of every chunk to a peer of class $C4$. This is because these peers have not enough capacity to distribute enough copies before new chunks are injected in the system, increasing the chances that new chunks inhibit the diffusion of the old ones. All the other policies can provide similar miss ratios.

\begin{figure}[t] 
\begin{center}
\subfigure[Class $C1$]{\includegraphics[width=0.4\textwidth]{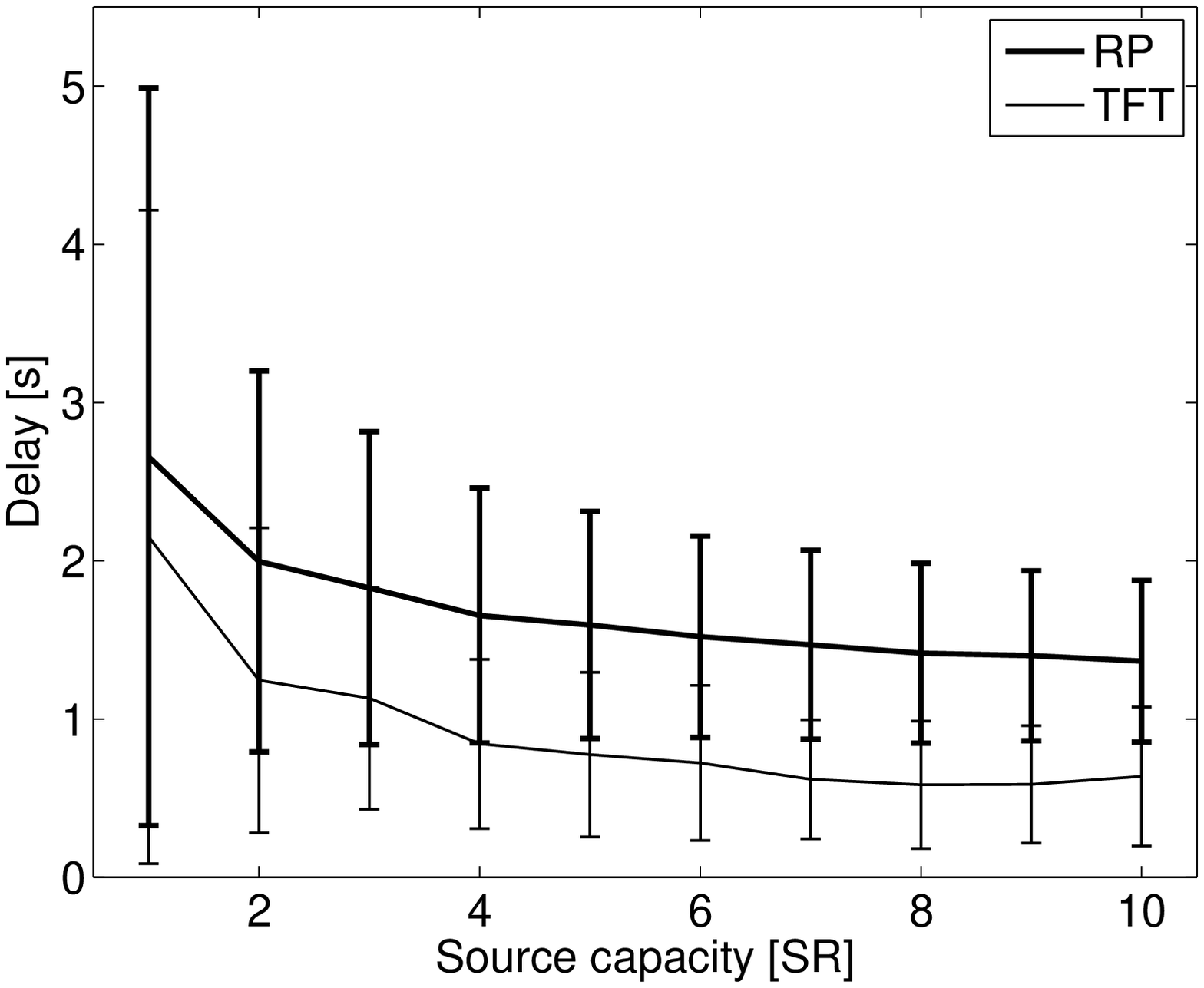}}
\subfigure[Class $C1$]{\includegraphics[width=0.4\textwidth]{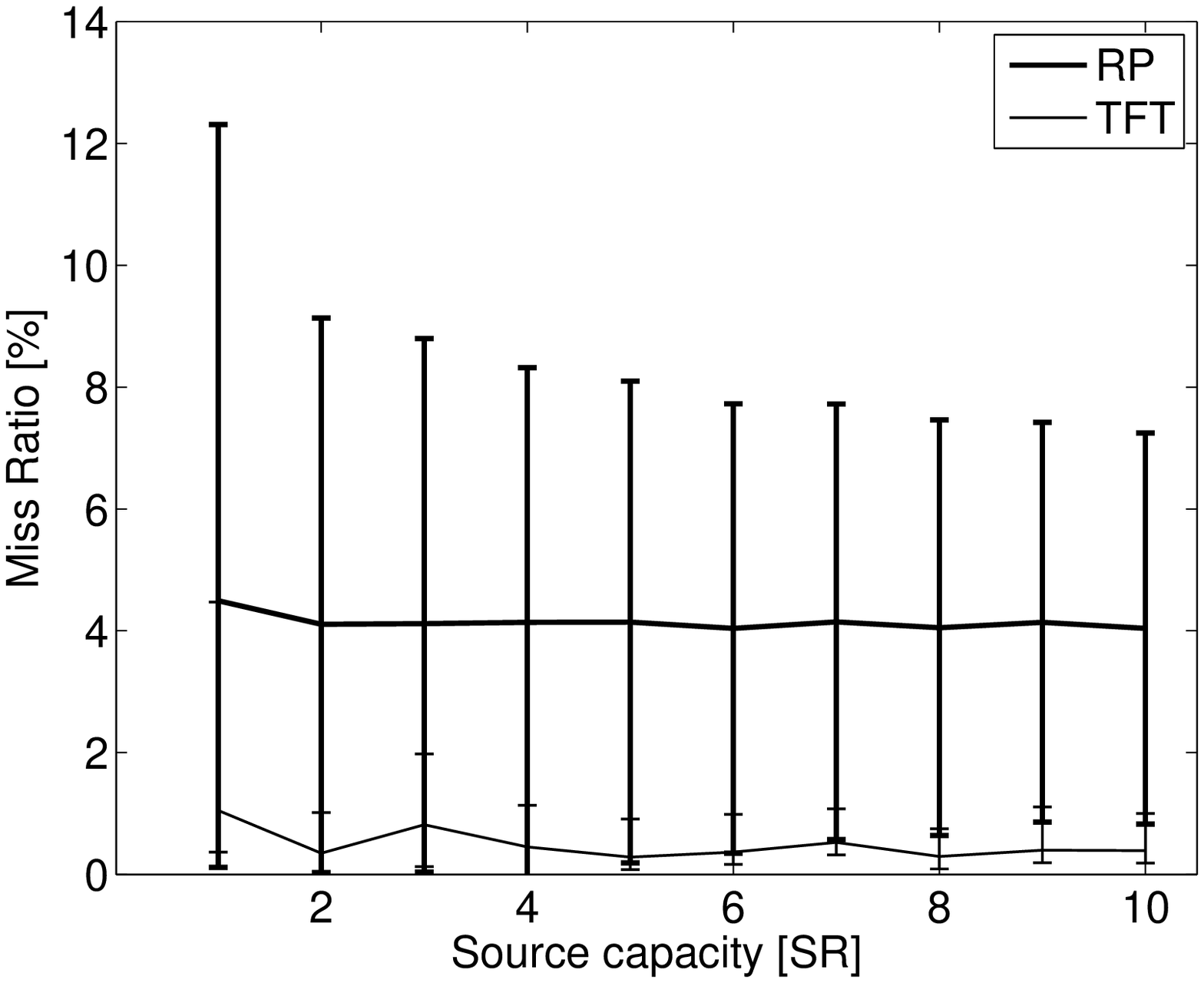}}\\
\subfigure[Class $C4$]{\includegraphics[width=0.4\textwidth]{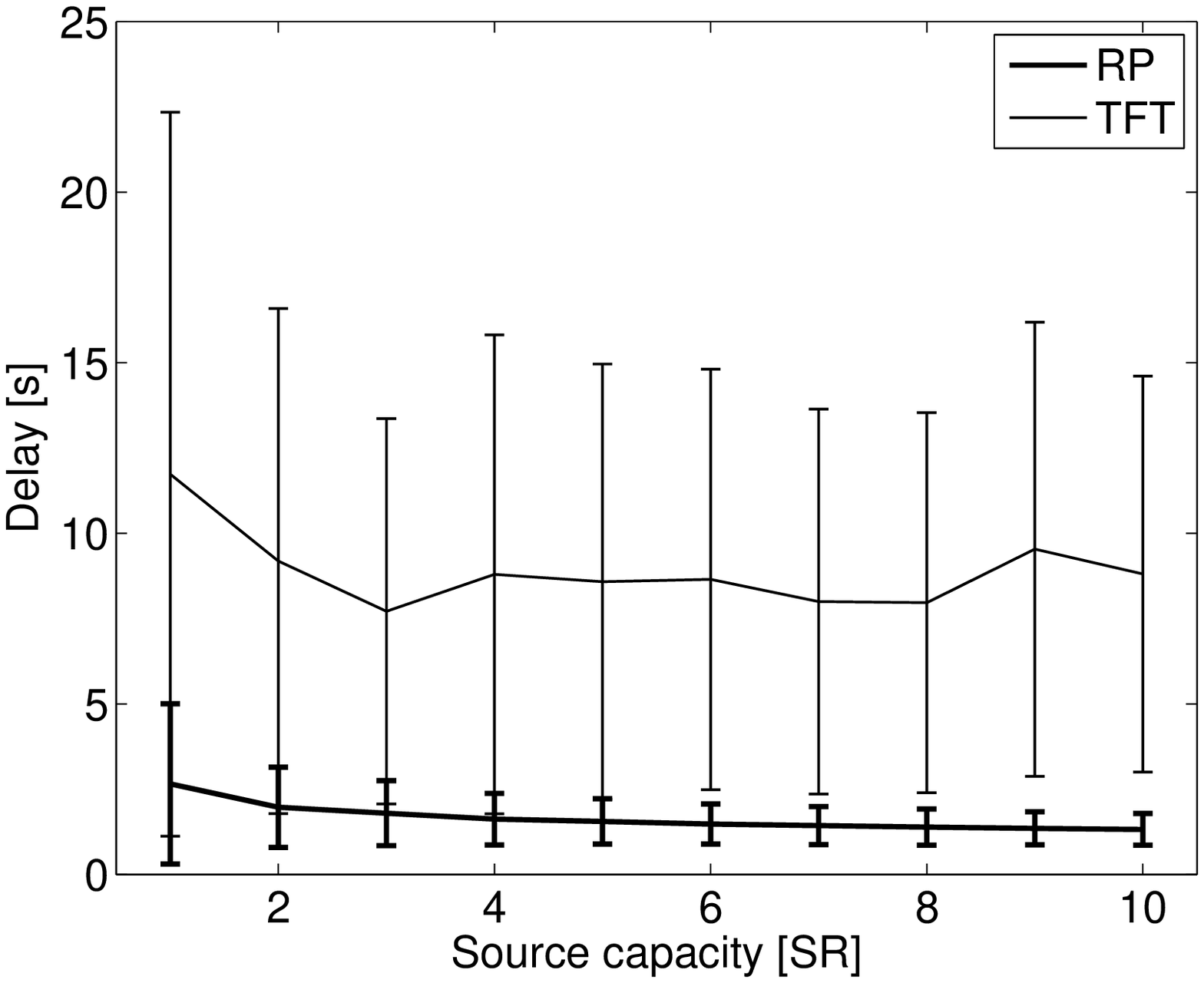}}
\subfigure[Class $C4$]{\includegraphics[width=0.4\textwidth]{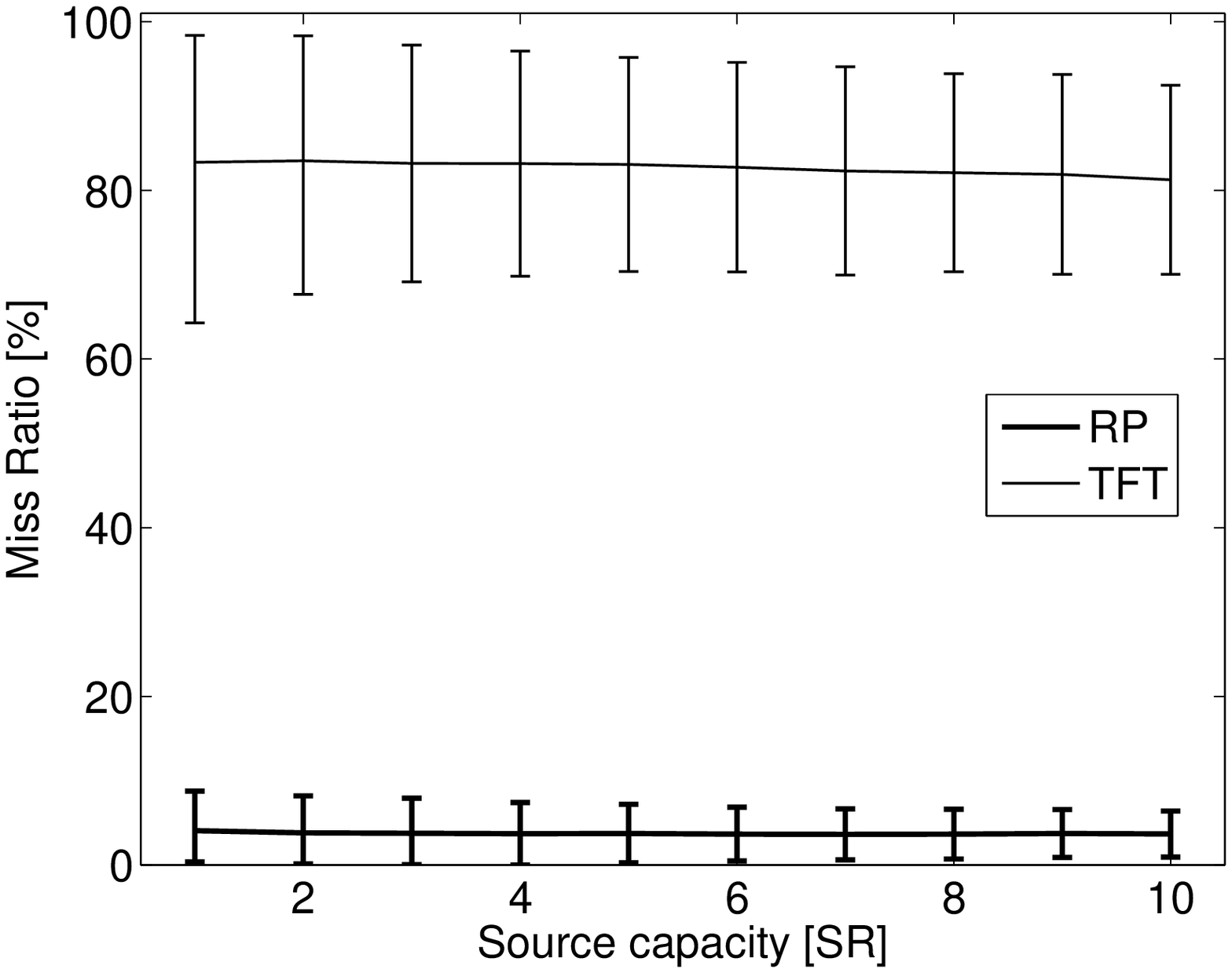}}
\caption{Diffusion delay and miss ratio (average value and its variance) as a function of the source upload capacity.}\label{fig:epidemics_aware_S2}
\end{center}
\end{figure}

We now investigate in more details the impact of the source upload capacity when it performs $RP$ selection. Results are reported in Figure~\ref{fig:epidemics_aware_S2} for $C1$ and $C4$. Nodes perform $RP$ or $TFT$ selection.  

The diffusion delay decreases as the number of copies of each chunk injected by the source increases. The decrease is particularly significant for the first additional copies ($u_s=2-3-4~SR$). This is because a chunk's initial diffusion tends to be exponential, so the delay improvement should be roughly proportional to the logarithm of the source capacity. For the miss ratio, we observe almost no gain by increasing the source capacity.

The variances of both the delay and miss ratio decrease by increasing the source upload capacity. Again, the first additional copies bring the larger variance decrease. This indicates the chunk diffusion is more stable, and schemes can provide steadier performance for the different chunks by increasing the source upload capacity.

\subsection{Convergence time and epoch length}
So far, we have highlighted that \emph{TFT} behaves similarly to \emph{BA} peer selection while being more appealing for real deployment. Such a scheme is driven by the evaluation of peer contributions performed every epoch $T_e$. As a consequence,  algorithms based on \emph{TFT} reach a steady-state where performance are stable after a certain period of time called \emph{convergence time}. 

\emph{TFT} convergence properties have already been analyzed for file-sharing applications in~\cite{gai.a.mathieu.f.ea:stratification}.
We investigate in this section the convergence time of \emph{TFT} peer selection in live streaming systems, and we evaluate the impact the epoch length $T_e$ has on their performance. 
In a live streaming system the \emph{convergence time}  indicates the time needed to reach both stable diffusion delay and miss ratio.


\begin{figure}[ht] 
\subfigure{\includegraphics[width=0.45\textwidth]{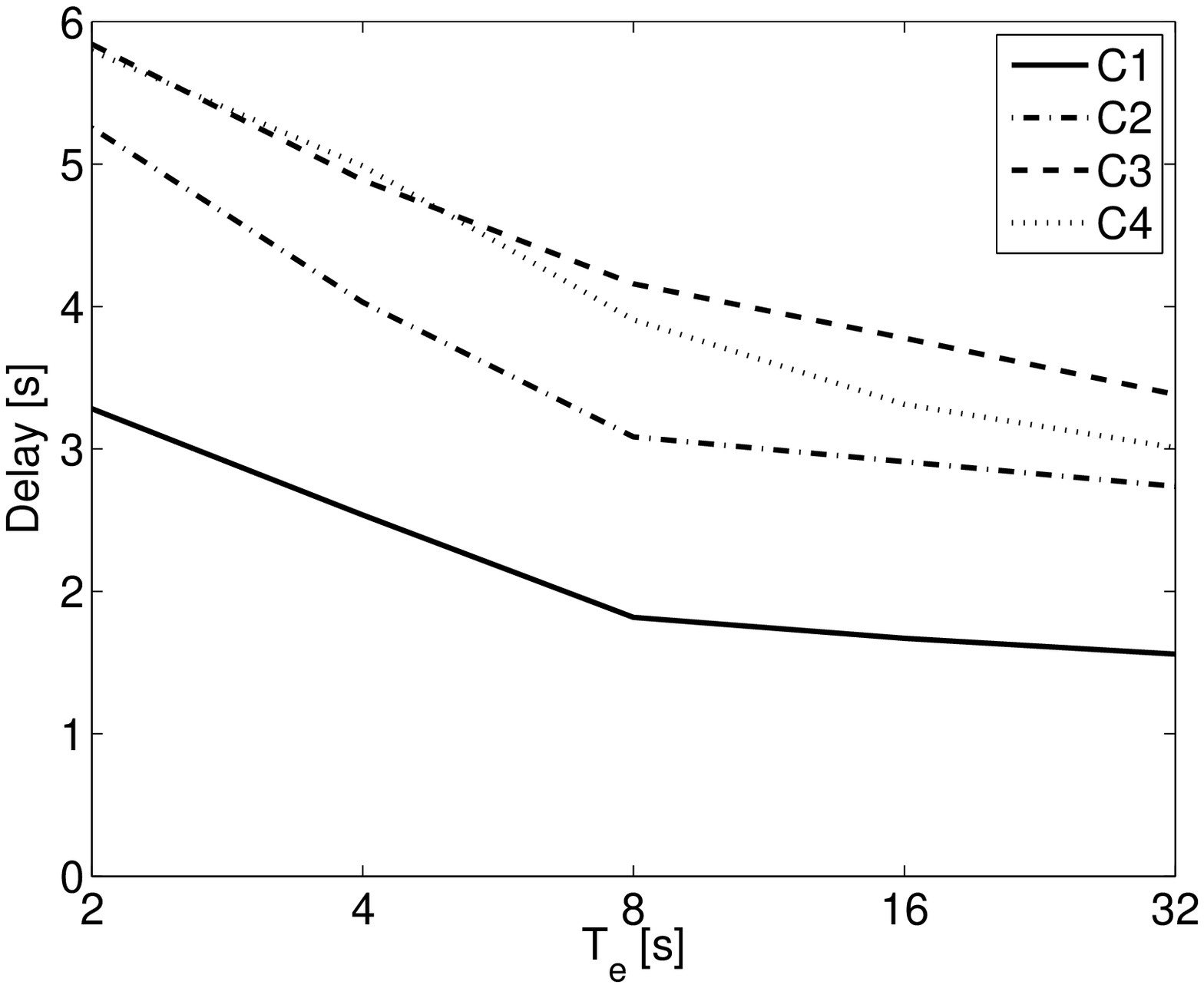}}
\subfigure{\includegraphics[width=0.45\textwidth]{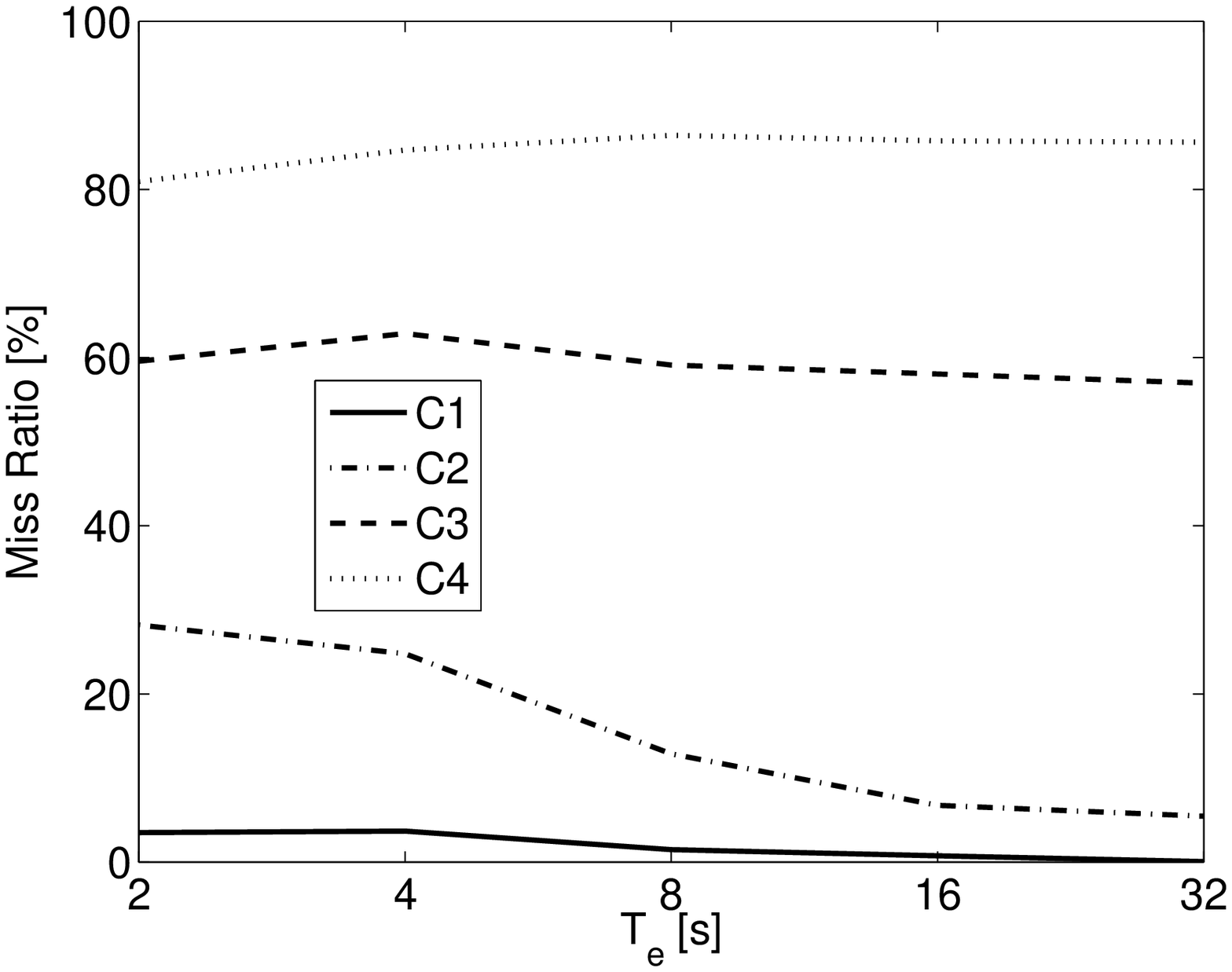}}\vspace{-0.2cm}
\caption{Diffusion delay and miss ratio as a function of the epoch length $T_e$.}
\label{fig:epidemics_aware_epoch}\vspace{-0.2cm}
\end{figure}

Figure~\ref{fig:epidemics_aware_epoch} indicates the diffusion delay decreases as the epoch length increases for all bandwidth classes. The miss ratio decreases as well only for richer classes, while for the poorer classes it stagnates or slightly increases. The larger evaluation time allows peers to better estimate the resources provided by their neighbors. As a consequence, the peer selection is more accurate and all peers improve their performance with respect to a $RP$ selection.

\begin{figure}[ht] 
\subfigure{\includegraphics[width=0.45\textwidth]{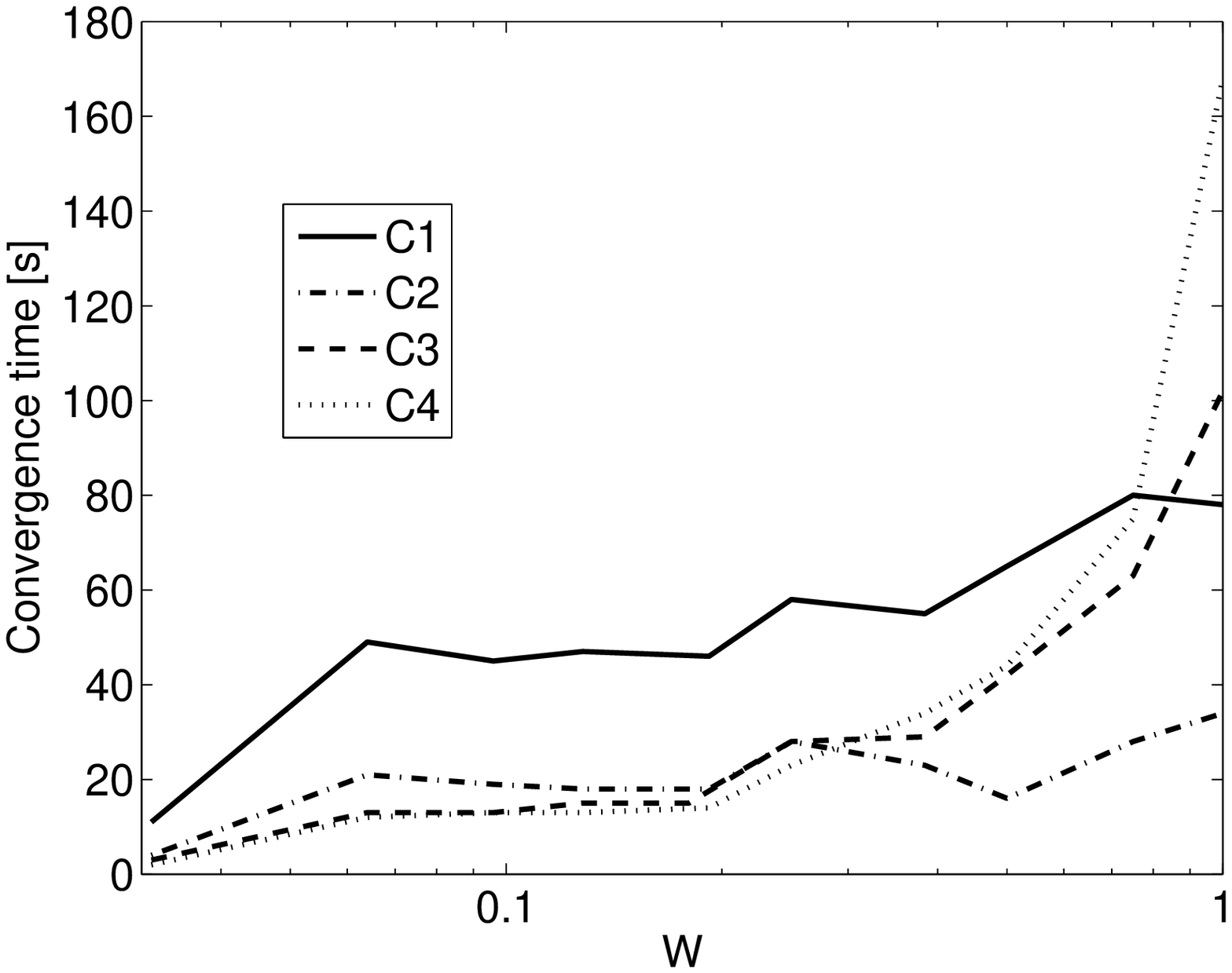}}
\subfigure{\includegraphics[width=0.45\textwidth]{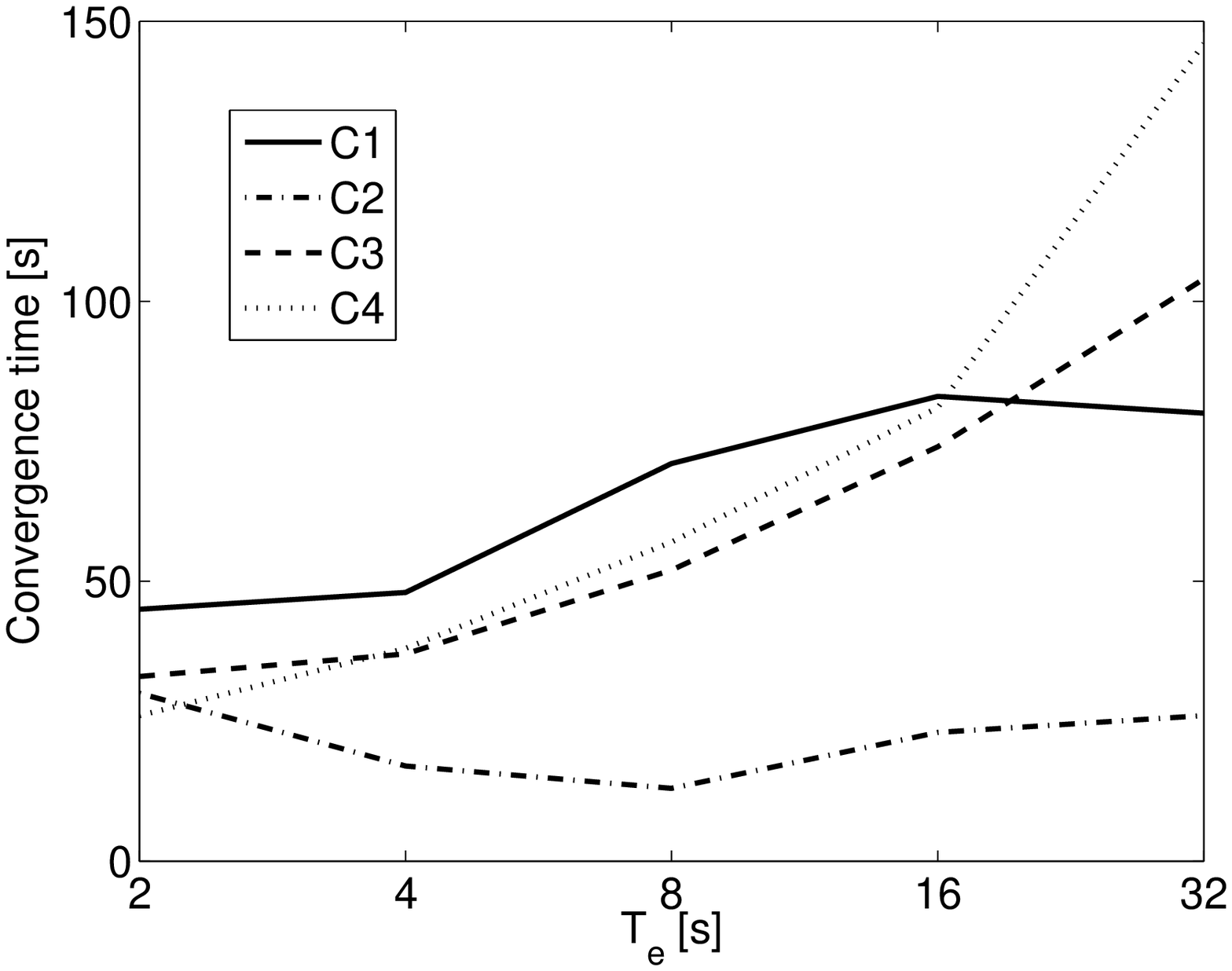}}\vspace{-0.3cm}
\caption{Convergence time as a function of the awareness probability for $T_e=10~s$, and of the epoch length for $W=0.75$.}\vspace{-0.5cm}
\label{fig:epidemics_aware_convergence}
\end{figure}

The price to pay is that longer epoch times require longer convergence times as showed in Figure~\ref{fig:epidemics_aware_convergence}. In details, peers of the richer classes require more time to reach a stable performance for small awareness parameters or short epoch lengths.
This because under such values only peers of the richer classes have performance different from $RP$ selection.
On the contrary, when $W$  or $T_e$ increases, the convergence time of poorer classes strongly increases. In such a case, the performance of the poorer classes is also affected, and, as a consequence, their convergence time increases and is eventually longer than the one of the richer classes.

\section{Conclusion}
In this paper, we have considered chunk distribution algorithms for unstructured peer-to-peer live streaming systems.

We have identified the first chunk exchanges as a key issue of the chunk diffusion process in heterogeneous systems. We have described some schemes designed to be aware of the resources shared by nodes, and we have provided a unified model to describe the peer selection of resource aware algorithms.
We have provided recursive formulas for the diffusion function of a generic \emph{resource aware peer/latest blind chunk selection} and validate their accuracy by means of simulations.

We have studied the performance of \emph{resource aware peer/ latest useful chunk} policies and we have shown that there exists a minimum value of resource awareness needed to improve the performance with respect to a random peer selection policy. We have highlighted a trade off between the performance of peers with different resources arising as a function of the level of awareness, and the strong impact that the source selection policy has on the diffusion process.


\newpage

\section*{Acknowledgment}

This work has been supported by the Collaborative Research Contract Mardi II between INRIA and Orange Labs, and by the European Commission through the NAPA-WINE Project, ICT Call 1 FP7-ICT-2007-1, Grant Agreement no.: 214412.

\newpage

\bibliographystyle{abbrv}
\bibliography{RR-7031}

\end{document}